\documentclass[10pt]{article}
\usepackage{graphicx}   
\usepackage{subcaption} 
\usepackage[toc]{appendix} 
\usepackage[utf8]{inputenc} 
\usepackage[english]{babel} 
\usepackage{geometry} 
\usepackage{graphicx} 
\usepackage{graphicx}
\usepackage{subcaption} 
\usepackage{changepage} 
\usepackage{float} 
\usepackage{longtable} 
\usepackage{listings} 
\usepackage{xcolor} 
\usepackage{enumitem} 
\usepackage{multirow} 
\usepackage{pdflscape} 
\usepackage{moresize} 
\usepackage{amsmath, amsthm, amssymb} 
\usepackage{algorithm} 
\usepackage{algpseudocode} 
\usepackage{array} 
\usepackage{changepage} 
\usepackage{booktabs} 
\usepackage{pifont}
\usepackage{url} 
\usepackage{diagbox}
\usepackage[colorlinks=true, citecolor=blue, linkcolor=blue, urlcolor=blue]{hyperref} 
\usepackage{makecell}
\usepackage{tabularx}
\usepackage{amsmath, amssymb}
\usepackage{tcolorbox}
\usepackage[table]{xcolor}
\usepackage{diagbox}
\usepackage{float}
\definecolor{lightgreen}{RGB}{235, 255, 235}
\usepackage{amsmath,amssymb,booktabs, makecell,hyperref,diagbox,float,enumitem,caption,graphicx }
\usepackage{pdflscape}              
\usepackage{longtable}              
\usepackage{booktabs}               
\usepackage{makecell}               
\usepackage{array}                  
\usepackage{hyperref}               
\usepackage{setspace} 
\def\BibTeX{{\rm B\kern-.05em{\sc i\kern-.025em b}\kern-.08em
    T\kern-.1667em\lower.7ex\hbox{E}\kern-.125emX}}
\usepackage{titlesec}
\titleformat{\section}
  {\normalfont\Large\bfseries}
  {\thesection}{1em}{}

\titleformat{\subsection}
  {\normalfont\large\bfseries}
  {\thesubsection}{1em}{}

\titleformat{\subsubsection}
  {\normalfont\normalsize\bfseries}
  {\thesubsubsection}{1em}{}
\begin{document}

\title{Analysing Skill Predominance in Generalized Fantasy Cricket
}
\author{
\textbf{Supratim Das}\\
\small \textit{Indian Statistical Institute, Kolkata}\\
\and
\textbf{Sarthak Sarkar}\\
\small \textit{Indian Statistical Institute, Kolkata}\\
\and
\textbf{Subhamoy Maitra}\\
\small \textit{Applied Statistics Unit}\\
\small \textit{Indian Statistical Institute, Kolkata}\\
\and
\textbf{Tridib Mukherjee}\\
\small \textit{Chief Data Scientist \& AI Officer}\\ \small \textit{IDfy, India}
}
\date{}
\maketitle

\begin{abstract}
In fantasy sports, strategic thinking-not mere luck-often defines who wins and who falls short. 
As fantasy cricket grows in popularity across India, understanding whether success stems from skill or chance has become both an analytical and regulatory question. 
This study introduces a new limited-selection contest framework in which participants choose from four expert-designed teams and share prizes based on the highest cumulative score. 
By combining simulation experiments with real performance data from the 2024 Indian Premier League (IPL), we evaluate whether measurable skill emerges within this structure. 
Results reveal that strategic and informed team selection consistently outperforms random choice, underscoring a clear skill advantage that persists despite stochastic variability. 
The analysis quantifies how team composition, inter-team correlation, and participant behaviour jointly influence winning probabilities, highlighting configurations where skill becomes statistically dominant. 
These findings provide actionable insights for players seeking to maximise returns through strategy and for platform designers aiming to develop fair, transparent, and engaging skill-based gaming ecosystems that balance competition with regulatory compliance.
\end{abstract}


\section{Introduction}

Fantasy sports have evolved into one of the most dynamic intersections of data analytics, behavioural strategy, and digital entertainment. 
What began as casual fan engagement has matured into a structured and competitive ecosystem where analytical reasoning, probabilistic assessment, and strategic decision-making determine success. 
Within this landscape, fantasy cricket occupies a particularly prominent position in India, fuelled by the sport’s deep cultural following and the accessibility of mobile-based real-money gaming platforms. 
The widespread participation in such contests has transformed cricket from a spectator activity into an arena for analytical skill and prediction, raising fundamental questions about the extent to which success depends on skill rather than luck.

The increasing complexity of fantasy sports has also attracted academic attention across economics, game theory, and computational social science. 
Early behavioural studies such as Farquhar and Meeds~\cite{b5} identified the motivational and cognitive drivers of fantasy participation, laying the groundwork for later analytical inquiries into strategic play. 
As the industry grew, researchers sought to quantify the relative influence of skill and chance on outcomes. 
A foundational contribution in this direction came from Getty et~al.~\cite{b1}, who developed a statistical framework to separate the effects of skill from randomness by analysing outcome variance across repeated contests. 
Their work introduced the measurable ``skill–luck continuum'' that has since guided much of the subsequent literature.

Building on this foundation, Haugh and Singal~\cite{b4} examined daily fantasy sports contests using econometric approaches to detect persistence in player performance, demonstrating that consistent success is unlikely to arise from chance alone. 
Duersch~\cite{b8} advanced the theoretical side by proposing formal metrics to distinguish skill from luck in probabilistic contests, arguing that even chance-heavy games retain identifiable skill components. 
Complementary empirical evidence was offered by O’Brien, Gleeson, and O’Sullivan~\cite{b7}, who analysed longitudinal data from the Fantasy Premier League to reveal consistent skill expression among top performers despite substantial stochastic variation. 
Collectively, these studies shifted the debate from a binary ``skill versus chance'' framing toward a continuum shaped by contest structure, information asymmetry, and participant heterogeneity.

Parallel to these analytical efforts, applied and policy-oriented research began exploring practical implications. 
The IIMB--FIFS report~\cite{b3} highlighted that the ability to measure and reproduce skill metrics is central to ensuring fair classification of contests under legal frameworks. 
Aishvarya, Das, and Kumar~\cite{b6} extended this argument by developing a decision-support framework that quantitatively evaluates the role of strategy in daily fantasy sports using operational data, bridging theoretical models and platform design. 
Meanwhile, Wilkins~\cite{b9} offered a bibliometric synthesis that mapped the rapid diversification of fantasy sports research, underscoring its evolution into a legitimate and data-driven field of inquiry. 
On the optimisation front, Singla and Shukla~\cite{b2} demonstrated how integer programming could enhance fantasy cricket team selection, reflecting the increasing fusion of computational methods with strategic gameplay.

Despite these advances, several research gaps persist. 
Most existing studies focus on open-selection formats where users build full teams from large player pools. 
However, emerging platforms increasingly employ simplified, \emph{limited-selection} formats, where users choose from a small number of expert-designed teams rather than constructing their own. 
This shift lowers cognitive entry barriers while retaining competitive depth, yet it remains underexplored academically. 
The strategic challenge in such contexts moves from player-level optimisation to meta-level evaluation—selecting among expert teams that embody distinct strategic logics.

Moreover, modern fantasy formats are becoming dynamically adaptive. 
In cricket, the introduction of the \emph{Impact Player} rule in the Indian Premier League (IPL) exemplifies a structural innovation that reshapes outcome distributions and introduces new layers of strategic uncertainty. 
Understanding whether such mechanisms amplify or dilute the influence of skill requires extending existing theoretical frameworks. 
Finally, limited integration of simulation-based models with real-world performance data has constrained the empirical validation of these insights.

Taken together, the literature establishes that skill is measurable, persistent, and structurally mediated. 
Yet its manifestation within constrained, evolving contest architectures—particularly those featuring limited-selection and dynamic tactical rules—remains an open question. 
The present study extends this conversation by analysing limited-selection fantasy cricket contests inspired by expert-team architectures and contemporary gameplay dynamics. 
By integrating simulation and empirical evaluation, it contributes to a deeper theoretical and operational understanding of how skill manifests and can be measured in modern fantasy sports environments.

\textbf{Plan of the Paper}

The remainder of this paper is organised as follows. 
Section~\ref{sec:theo_basic} introduces the conceptual foundation and the baseline simulation framework that define the proposed contest structure. 
It outlines the core assumptions, the setup of analytical and random participants, and the mechanisms through which contest outcomes and skill visibility are generated. 
Sections~\ref{sec:theo_corr} and~\ref{sec:theo_impact} then extend this framework along two important directions. 
Section~\ref{sec:theo_corr} examines how varying the number of experts and the degree of correlation among their performances influences the strategic depth of the contest, while Section~\ref{sec:theo_impact} incorporates an \emph{Impact Player} feature to evaluate whether this additional layer of decision-making enhances the skill component of participation. 

Section~\ref{sec:real} transitions to the empirical analysis using IPL~2024 data, testing the hypotheses developed in the theoretical sections under real-world conditions. 
By varying the number of common players and expert configurations, this section illustrates how platform design parameters shape observed performance differences and the extent of skill expression. 
Section~\ref{sec:real_impact} further extends this analysis by integrating the \emph{Impact Player} mechanism into real match settings, allowing direct comparison with the simulation-based insights. 

Section~\ref{sec:regression} presents regression-based analyses to quantify the relationships identified earlier and assess their statistical significance. 
Finally, Section~\ref{sec:conclusion} summarises the key findings and discusses their broader implications for contest design, participant behaviour, and the understanding of skill in fantasy sports.

\section{A Simulation Based Analysis}
\label{sec:theo_basic}
\subsection{Basic Definitions and Key Assumptions}
We begin by introducing some \textbf{basic notation and assumptions} regarding the \textbf{distribution of scores among the expert teams}. Let the number of expert teams be denoted by $n$, and let $P_1, P_2, \dots, P_n$ represent the points scored by these teams in a given contest. In order to capture the inherent variability and correlation in team performances, we assume that the joint distribution of the scores follows a multivariate lognormal distribution:

\[
(P_1, \dots, P_n) \sim \text{LogNormal}(\boldsymbol{\mu}, \boldsymbol{\Sigma})\
\]

where $\boldsymbol{\mu}$ is the mean vector and $\boldsymbol{\Sigma}$ is the covariance matrix of the associated multivariate normal distribution. Equivalently, when we take logarithms of the scores, we obtain a multivariate normal structure:
\[
(\log P_1, \dots, \log P_n) \sim \mathcal{N}(\boldsymbol{\mu}, \boldsymbol{\Sigma})
\]

This formulation is appealing, as it allows the log-scores to be modeled within a \textbf{Gaussian} framework, while ensuring that the original scores are strictly positive, as required in a fantasy contest setting. For analytical tractability, and to reduce the number of free parameters, we further assume that the correlation structure across teams is equicorrelated with a common correlation parameter $\rho = 0.4$. Thus, any two teams’ scores share a moderate dependence.

\noindent Next, we turn to the problem of determining the \textbf{win probability of each team}. Let $\pi_i$ denote the probability that the $i^\text{th}$ expert team achieves the maximum score among all competing teams. Formally, this is expressed as:
\[
\pi_i = \mathbb{P}(P_i \ge P_j \quad \forall j \neq i).
\]

In other words, $\pi_i$ is the probability that team $i$ is the unique winner of the contest, given the joint distribution of the log-scores. While this definition is conceptually straightforward, the actual computation of these probabilities is analytically challenging due to the dependence structure induced by the covariance matrix $\boldsymbol{\Sigma}$. Closed-form expressions are generally unavailable, and exact evaluation would require integration over a high-dimensional region of the multivariate Gaussian distribution. To overcome this difficulty, we employ a simulation-based approach, wherein we draw repeated samples from the distribution of $(P_1, \dots, P_n)$ and estimate $\pi_i$ empirically as the frequency with which team $i$ emerges as the winner. This provides a flexible and practical method for approximating the true win probabilities.

\noindent For our simulation analysis, we  introduce some behavioral assumptions regarding the players participating in the contest. Broadly, we classify them into two categories based on their decision-making sophistication and access to information: \textbf{analytical players} and \textbf{random players}.

An \textbf{analytical player} is characterized by their ability to form probabilistic estimates of the true win probabilities, denoted by $\boldsymbol{\pi} = (\pi_1, \dots, \pi_n)$. Rather than having perfect information, such a player constructs beliefs about these probabilities by sampling from a Dirichlet distribution,
\[
\mathbf{p} \sim \mathrm{Dirichlet}(\alpha \boldsymbol{\pi}),
\]
where $\alpha > 0$ serves as a concentration parameter that controls the accuracy and confidence of these estimates. A higher value of $\alpha$ corresponds to a more concentrated distribution around the true probabilities, thereby representing a more precise and confident player. To calibrate this parameter, we impose the requirement that the sampled estimates $\mathbf{p}$ lie within a tolerance bound $\beta$ of the true probabilities with high probability. Using Chebyshev’s inequality in conjunction with the union bound (see Appendix~\ref{app:dirichlet}), it can be shown that a sufficient condition for this guarantee is
\[
\alpha \ge \frac{n}{4 \delta \beta^2} - 1,
\]
where $\delta$ denotes the allowable probability of violation. By choosing $\alpha$ according to this condition, we ensure that analytical players generate estimates that are reliably close to the true probabilities. Once the probability vector $\mathbf{p}$ has been obtained, the team choice of an analytical player is made by sampling according to $\mathbf{p}$:
\[
\text{Team choice} \sim \mathrm{Multinomial}(1, \mathbf{p}).
\]
In essence, the analytical player behaves as though they have access to noisy but well-calibrated estimates of the underlying win probabilities.

In contrast, a \textbf{random player} does not attempt to form any such estimates or reason about underlying win probabilities. Their decision-making is entirely uninformed and uniformly distributed across all possible teams. Concretely, the random player chooses a team according to
\[
\text{Team choice} \sim \mathrm{Multinomial}\Big(1, \frac{\mathbf{1}_n}{n}\Big),
\]
where $\mathbf{1}_n$ denotes the vector of all ones of length $n$. This represents a purely chance-driven strategy, independent of any performance-based information or analytical reasoning. \\

\noindent The following example demonstrates the contest gameplay and prize money distribution. Suppose a total of $N$ players enter the contest, each paying a fixed entry fee of Rs.~25. As before, let $\tau$ denote the fraction of \textbf{analytical players} and $(1-\tau)$ the fraction of \textbf{random players}. For each expert team $i$, the number of players choosing that team is given by
\[
N_i = N_i^\tau + N_i^{1-\tau},
\]
where $N_i^\tau$ and $N_i^{1-\tau}$ denote the number of analytical and random players selecting team $i$, respectively. This formulation allows us to capture heterogeneity in player behavior within the same contest environment.

The payoff mechanism considered here follows a \textbf{shared prize pool} structure, which is the standard approach employed by most fantasy gaming platforms. All entry fees collected from participants form a common pool, from which a fixed percentage is retained by the platform as commission. For instance, with $N = 1000$ players, an entry fee of Rs.~25 per player, and a $20\%$ platform cut, the total prize pool available for distribution becomes
\[
\text{Prize pool} = \frac{25 \times 1000 \times (100 - 20)}{100} = \text{Rs. 20000}.
\]

If the $i^{\text{th}}$ expert team wins, all $N_i$ players who selected that team share the prize pool equally. The payoff for the $k^{\text{th}}$ player is therefore
\[
W_k =
\begin{cases}
\frac{20000}{N_i}, & \text{if the $k^{\text{th}}$ player selected team $i$},\\[6pt]
0, & \text{otherwise}.
\end{cases}
\]
This setup naturally induces a strategic trade-off: selecting a stronger team increases the probability of winning, but if many players converge on the same team, the resulting payoff per player is diluted. From the platform’s perspective, this structure is advantageous since the total payout is capped at the fixed prize pool, minimizing financial exposure.

Overall, the shared prize pool mechanism encourages players to balance \textbf{winning likelihood} against \textbf{crowding risk}. Analytical players may attempt to forecast popular team selections to optimize expected payoffs, while random players contribute to unpredictability in the distribution of selections. Together, these dynamics make the contest environment both competitive and stochastic, aligning closely with the incentive structure observed in real-world fantasy gaming platforms.
\subsection{Various Simulation Configurations}

In order to systematically study the performance of players under different conditions, we define a set of simulation parameters and configurations. Our simulations focus on a contest involving four expert teams and a population of players with heterogeneous strategic behavior. For each configuration, the true win probabilities of the teams are estimated by drawing 100,000 samples from the corresponding multivariate normal distribution of log-scores. Once these probabilities are established, we simulate the matches themselves 10,000 times for each configuration. Therefore, the total number of iterations per configuration is 10,000, providing a robust basis for statistical analysis of player winnings.\\

\noindent All simulation runs are conducted under a shared set of baseline parameters. Each contest involves $n = 4$ expert teams and a total of $N = 1000$ participants. A fraction $\tau$ of these participants are \textit{analytical players}, who attempt to estimate the true win probabilities based on the known distribution, while the remaining $1000(1-\tau)$ participants choose teams randomly; initially, $\tau$ is set to $0.2$. Each player contributes an entry fee of Rs.~$25$, from which the platform deducts a $20\%$ commission, allocating the remaining $80\%$ to the prize pool for distribution among the winners. The log-transformed scores of the expert teams are assumed to have a moderate pairwise correlation of $0.4$, capturing the interdependence in their performances. Finally, analytical players’ estimated win probabilities are assumed to lie within an error bound of $\pm 0.04$ of the true probabilities with a 95\% confidence level, reflecting a realistic degree of estimation accuracy.\\

\noindent The simulation study is performed over three distinct configurations, each designed to highlight different aspects of competitive dynamics and variability in team performance. Table~\ref{tab:configurations} summarizes the parameters of these configurations:


\begin{table}[h]
\centering
\begin{tabularx}{\textwidth}{@{} l c c X @{}}
\toprule
\textbf{Name} & \textbf{Mean Points} & \textbf{Std. Deviations} & \textbf{Interpretation} \\
\midrule
Equi-mean\_Equivariance & $[500, 500, 500, 500]$ & $[60, 60, 60, 60]$ & All teams are statistically identical in both mean and variance, serving as a fairness benchmark. \\
\addlinespace
Unequal-mean\_Equivariance & $[440, 473, 517, 550]$ & $[60, 60, 60, 60]$ & Teams differ in average performance (mean) but maintain equal variability (standard deviation). \\
\addlinespace
Unequal-mean\_Unequal-std & $[440, 473, 517, 550]$ & $[60, 30, 60, 30]$ & A realistic asymmetric setting where teams vary both in average strength and consistency, reflecting diverse competitive environments. \\
\bottomrule
\end{tabularx}
\caption{Simulation Configurations for Expert Teams}
\label{tab:configurations}
\end{table}

\medskip

For each configuration, our analysis focuses on the \textbf{expected winnings of an individual player}, conditional on the team they select. Specifically, we compute the distribution of winnings for a player who chooses the $i^{th}$ expert team, for all $i \in \{1,2,3,4\}$, under both the shared prize pool and fixed-multiple payoff structures. This approach allows us to capture the effect of team strength, variability, and the composition of analytical versus random players on the potential outcomes for each participant.

\subsection{Key Metrics for Player Behavior}

To analyze player performance across the different configurations, we focus on two key metrics: the \emph{Selection Ratio} and the \emph{Mean Winnings} including zeros. These metrics help quantify both strategic decision-making and expected payoffs for players.

\noindent The \textbf{Selection Ratio} measures how often analytical players select a given expert team relative to random players. It is computed as the weighted ratio of analytical to random selections, normalized by the total number of analytical and random players in the tournament. A ratio greater than one indicates that analytical players are disproportionately choosing a particular team, suggesting that they are effectively identifying stronger teams.  

 \textbf{Example:} In a tournament with $N=1000$ players, suppose $400$ are analytical and $600$ are random. If $120$ analytical and $90$ random players select Expert Team 1, the Selection Ratio for Team 1 is calculated as

\[
\text{Selection Ratio}_1 = \frac{120/400}{90/600} = 2,
\]

\noindent indicating that analytical players are twice as likely to select Team 1 compared to random players.

The \textbf{Mean Winnings (including zeros)} represents the expected payoff for a player choosing a specific team, accounting for both winning and losing iterations. By including zeros from losing attempts, this metric reflects the true expected monetary outcome rather than only successful payouts.  

\textbf{Example:} Suppose Team 2 wins in $2,500$ out of $10,000$ simulated iterations, and each winning payout is Rs.~50. Then the mean winnings for a player selecting Team 2 are

\[
\text{Mean Winnings}_2 = \frac{2,500 \times 50 + 7,500 \times 0}{10,000} = \text{Rs.~12.5}.
\]

Together, these metrics provide insight into both player strategy (who chooses which teams) and the expected monetary outcomes under the simulation conditions, allowing for a comprehensive analysis of player behavior across different configurations.

\subsection{Simulation Results and Insights}
\begin{figure}[H]
    \centering
    \includegraphics[width=0.8\linewidth]{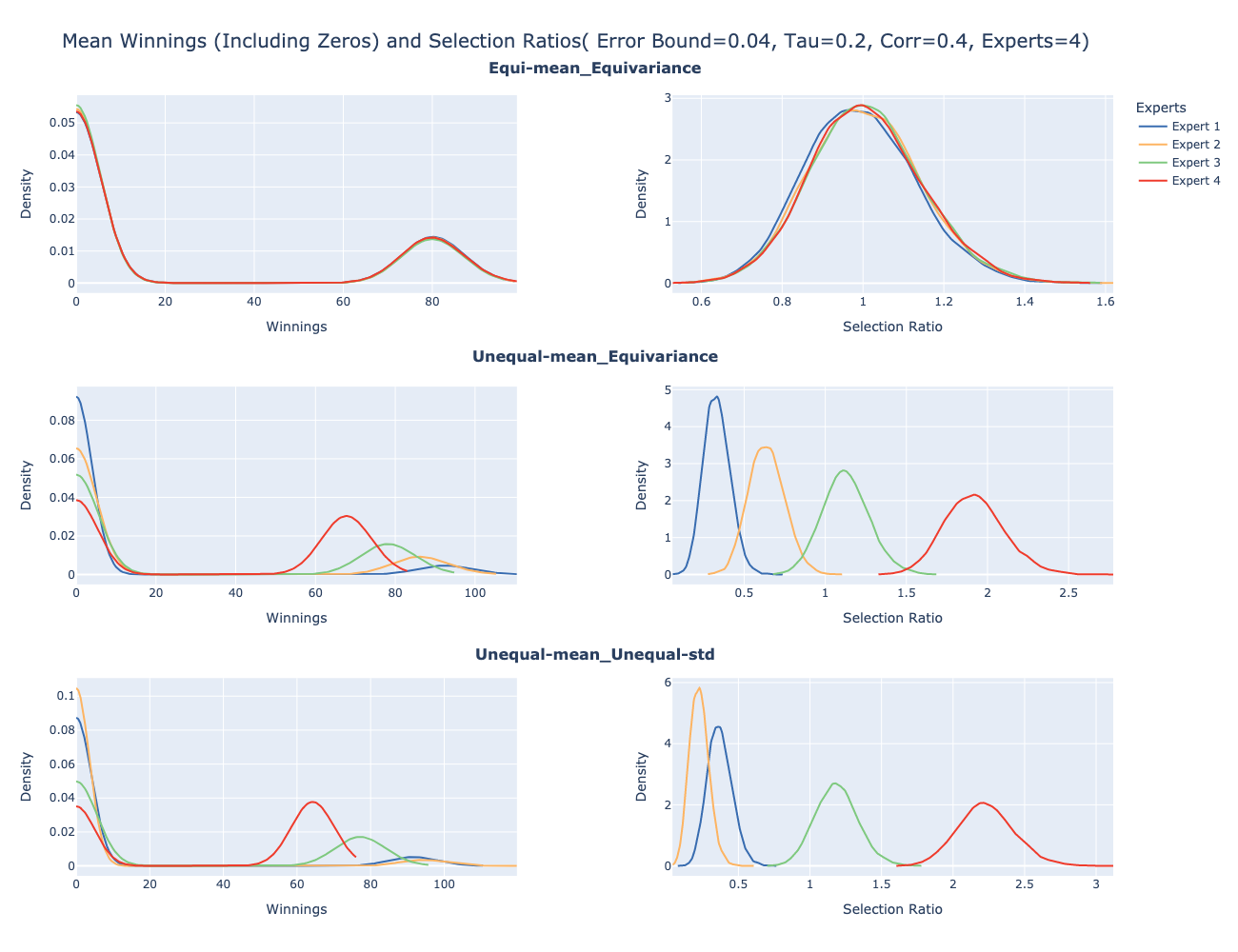}
    \caption{Summary Plots}
    \label{fig:summary}
\end{figure}
From \ref{fig:summary}, we can see that analytical players tend to select the better team ($4^{th}$ expert) on average, thus reflecting in high selection ratio for the $4^{th}$ expert team in the Unequal-mean\_Equivariance and Unequal-mean\_Unequal-std cases. However, where the experts are symmetric among themselves, the analytical players fail to grasp any extra advantage and play like random players.\\

\noindent Another key observation is that, if a player should choose the best team (Team 4), then he/she would, on an average, win less money. This is due to the reason that since more skillful players are selecting the best team, the prize amount is getting divided among a larger pool of people, thus reducing the reward of winning the contest.\\

Now, from these observations, we can see that the presence of more skillful players could reduce the skill premium( reward of winning). Also, the accuracy with which which the analytical players estimate the win probabilities could affect the quality of their estimates and in turn, winning in the contest.
So, it is important to study  how the fraction of analytical players ($\tau$) and the accuracy of their probability estimates (error bound $\beta$) affect performance. We focus on metrics for the best-performing team: \emph{mean winnings} (including zeros) and the weighted selection ratio for analytical versus random players.

Simulations are conducted for $\tau = 0.1, 0.2, 0.3, 0.4$ with fixed $\beta = 0.04$, to study the effect of varying the fraction of analytical players. We also vary $\beta$ from 0.02 to 0.05 with $\tau = 0.2$, to examine the influence of estimation accuracy.

\begin{figure}[H]
    \centering
    \includegraphics[width=0.65\linewidth]{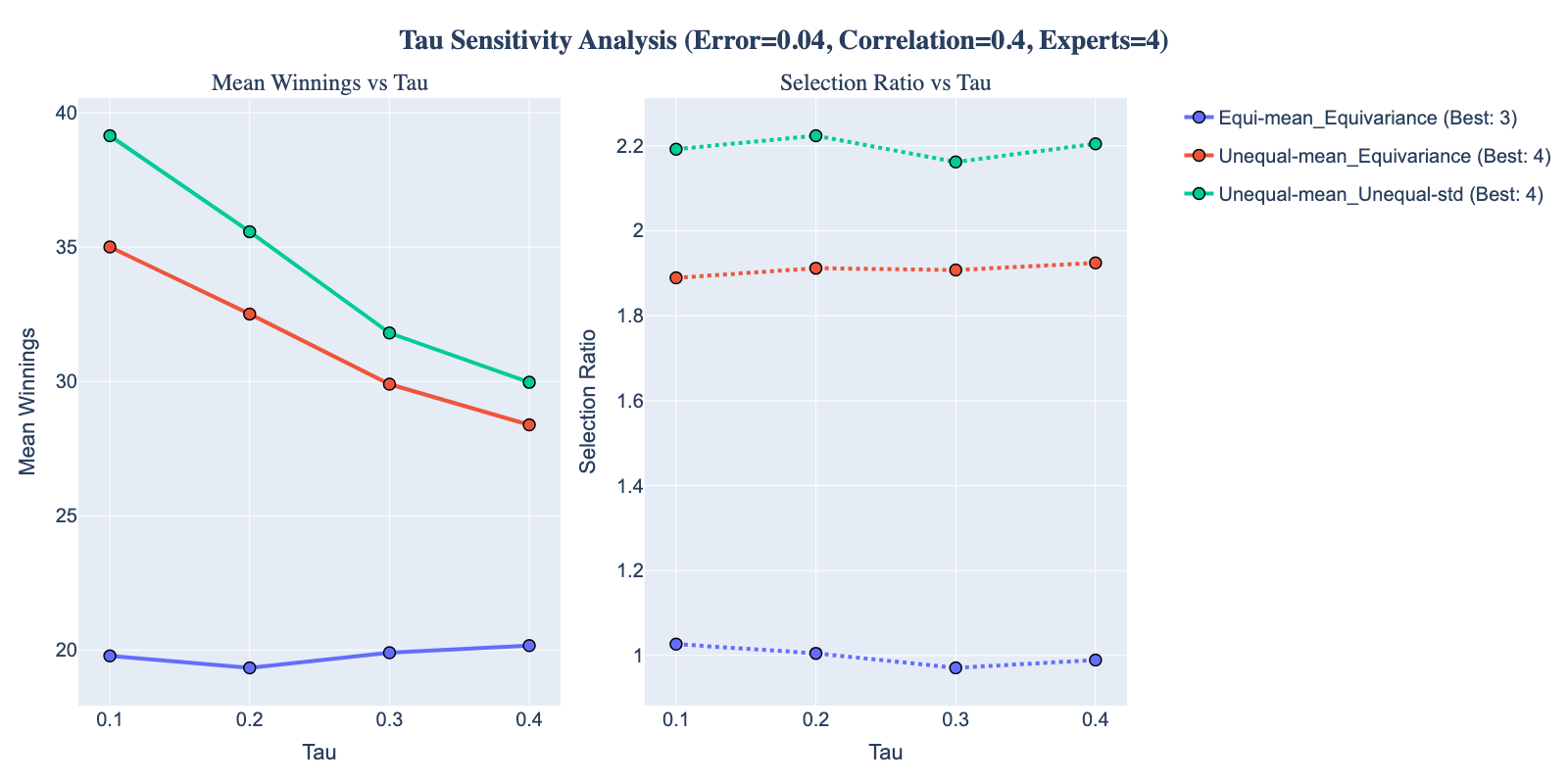}
    \caption{Mean winnings and selection ratio across $\tau$ (Error Bound $\beta=0.04$).}
    \label{fig:across_tau_single}
\end{figure}

\begin{figure}[H]
    \centering
    \includegraphics[width=0.65\linewidth]{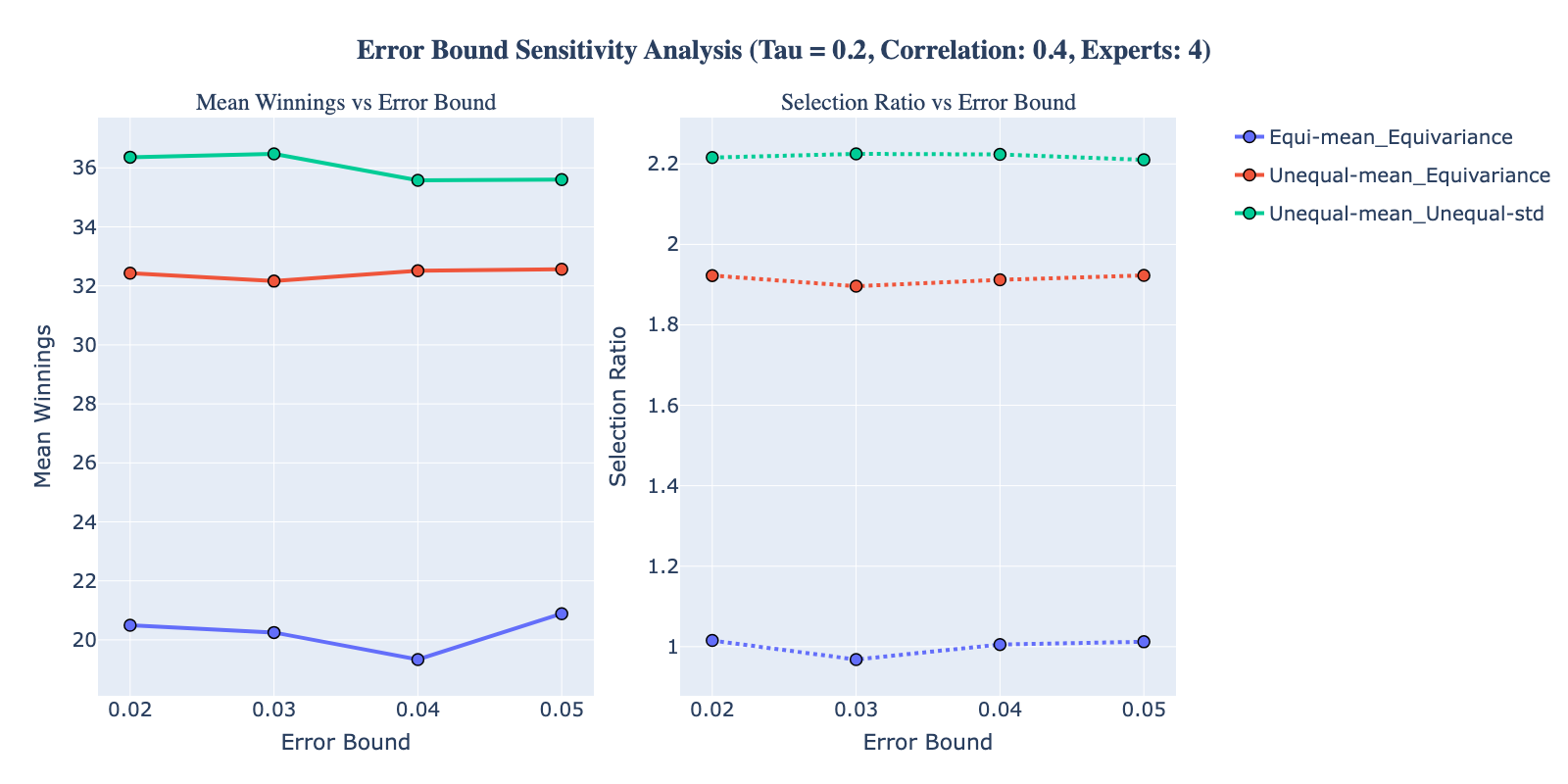}
    \caption{Mean winnings and selection ratio across $\beta$ (Fraction of Analytical Players $\tau=0.2$).}
    \label{fig:across_beta_single}
\end{figure}

Several clear patterns emerge from the simulations and visualizations.  

When \textbf{analytical players are few}, their \textbf{strategic advantage} is strongest. In such cases, the \emph{mean winnings} for those selecting the best team are noticeably higher, highlighting the benefit of \textbf{informed decision-making} in a largely random participant pool. As the proportion of analytical players increases, more participants make informed selections, which \textbf{dilutes individual advantage} and slightly reduces per-player payouts. 

\textbf{Estimation accuracy} also plays a key role. Players with more precise probability estimates (smaller $\beta$) generally achieve \textbf{higher mean winnings}, while less accurate estimates modestly reduce expected payouts. However, the effect of moderate estimation errors is relatively mild, indicating that \textbf{analytical players retain a competitive edge} even under some uncertainty.  

Taken together, these results suggest that the \textbf{optimal strategy} for an analytical participant involves balancing \textbf{scarcity} and \textbf{prediction accuracy}. Capitalizing on relative scarcity among analytical players can enhance expected rewards, while maintaining high-quality estimates ensures that \emph{strategic choices} are consistently aligned with the underlying probabilities. 

Finally, the simulations provide evidence of \textbf{skill} in this format. \textbf{Analytical players} are consistently more likely to select the \textbf{best-performing team} compared to random players, demonstrating that \textbf{rational decision-making} has a significant influence on outcomes in the contest.

\section{Extensions}
\label{extension}

So far, we have analyzed participant behavior and outcomes under a fixed setup of experts and correlation assumptions. However, in reality, the structure of the game can vary in important ways. Two of the most critical factors that can change the balance between \textbf{skill} and \textbf{luck} are: (i) the \textbf{number of experts} providing predictions, and (ii) the \textbf{correlation among the teams} constructed from those experts’ opinions. Both of these directly affect the diversity of choices available to participants, and hence, the strength of any \textbf{skill advantage}. This is analyzed in section \ref{sec:theo_corr}. 

An additional feature that can be incorporated into the game is the \textbf{Impact Player}. This feature allows the participant to engage more deeply in team selection strategy. Specifically, the participant begins by choosing one of the available expert teams. Following this, they are given the opportunity to select a single player from the remaining pool of cricket players in the match, who will serve as the "Impact Player" for that team. 
At the conclusion of the match, the points accumulated by the impact player are compared with those of the lowest-contributing player in the selected expert team. If the impact player's performance surpasses that of the lowest scorer, the contribution of the lowest-performing team member is replaced by the points of the impact player. This mechanism ensures that the user's team benefits directly from the strategic selection of the impact player. This is discussed in section \ref{sec:theo_impact}.

\subsection{Impact of Correlations and Number of Experts}
\label{sec:theo_corr}

The \textbf{number of experts} essentially determines the richness of the prediction pool. If there are very few experts, participants face only a small set of teams to choose from, which may reduce the scope for \textbf{analytical skill} to shine. In such a restricted environment, even a random participant may occasionally land on a strong team simply by chance.  

On the other hand, when there are too many experts, the situation can also become less favorable for \textbf{skill} to dominate. With a very large number of expert teams, overlaps in player selections become inevitable, and the marginal distinction between teams decreases. This increases the probability that many teams perform similarly, \textbf{diluting the benefit of careful decision-making}.  

Thus, it is important to study intermediate cases and identify an \textbf{optimal number of experts} where analytical decision-making confers a clear and consistent edge. For this purpose, we consider simulations with $2$, $3$, and $4$ experts, covering the range from a relatively simple choice set to a moderately complex one.

In real-world fantasy sports and prediction markets, the teams formed from expert opinions are not independent of one another. The same star players often appear across multiple teams, leading to a high degree of overlap. This overlap induces statistical \textbf{correlation}: the performance of one team is no longer independent of another.  

For example, if two teams share the same captain and several key batsmen, then both teams will rise or fall together depending on those players’ performances. This \textbf{correlation} reduces the variety of outcomes across the set of expert teams and makes it harder for participants to gain a significant \textbf{advantage} by carefully selecting one team over another.

To examine this systematically, we impose different correlation structures in our simulations: $\rho = 0.1, 0.2,...,0.9$. These values represent low to high correlation scenarios. By comparing across these, we can see how robust the \textbf{skill advantage} remains in environments where team performances are increasingly similar.

By comparing these two setups, we can measure how much extra \textbf{skill-based advantage} arises when participants are empowered to make leadership choices, rather than being constrained by expert defaults.

The simulations are therefore designed to cover the following dimensions:  
\begin{itemize}
    \item \textbf{Number of experts:} $2, 3, 4$  
    \item \textbf{Correlation levels:} $0.1, 0.2,...,0.9$  
\end{itemize}

As before, for each combination of these parameters, we track key outcome metrics, including \textbf{mean winnings} (with zeros included), and \textbf{selection ratios} for the best team.


\begin{figure}[h]
    \centering
    \begin{subfigure}[b]{0.45\textwidth}
        \centering
        \includegraphics[width=0.8\textwidth]{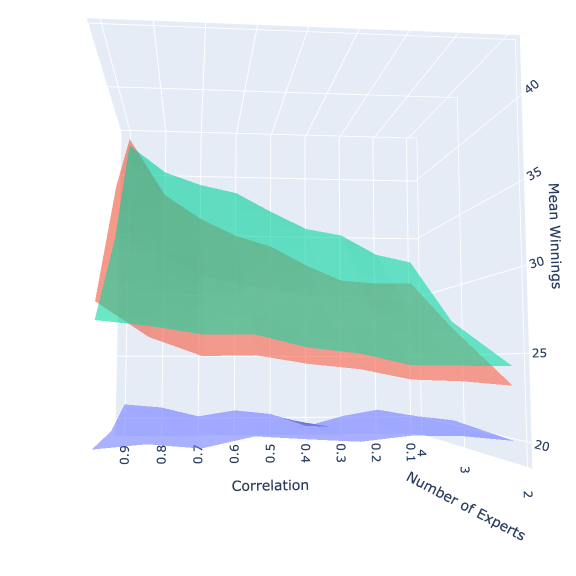}
        \caption{Across correlations}
        \label{fig:img1}
    \end{subfigure}
    \hfill
    \begin{subfigure}[b]{0.45\textwidth}
        \centering
        \includegraphics[width=0.8\textwidth]{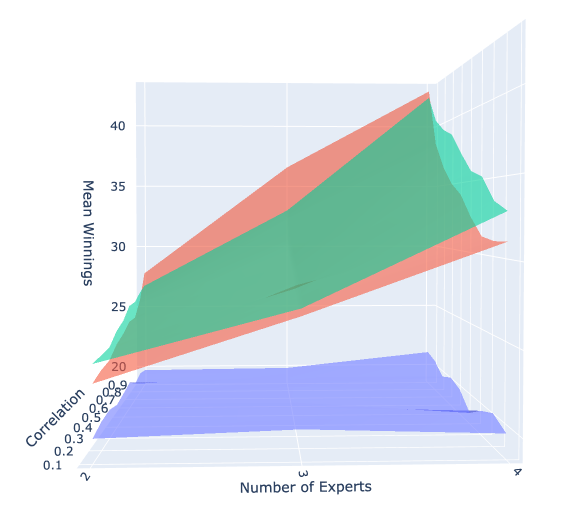}
        \caption{Across differing no. of experts}
        \label{fig:img2}
    \end{subfigure}
    
    \caption{Mean Winnings (with zeros included)}
    \label{fig:side_by_side1}
\end{figure}

\begin{figure}[h]
    \centering
    \begin{subfigure}[b]{0.49\textwidth}
        \centering
        \includegraphics[width=0.8\textwidth]{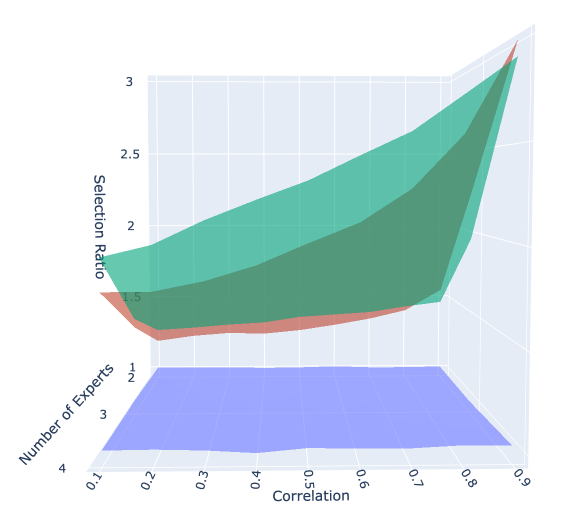}
        \caption{Across correlations}
        \label{fig:across_corrs}
    \end{subfigure}
    \hfill
    \begin{subfigure}[b]{0.49\textwidth}
        \centering
        \includegraphics[width=0.8\textwidth]{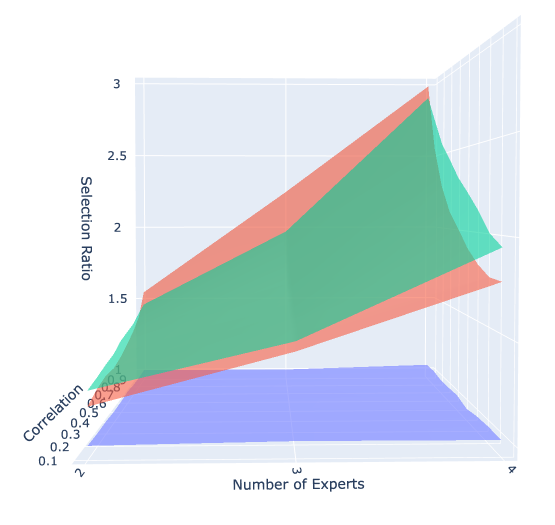}
        \caption{Across differing no. of experts}
        \label{fig:across_experts}
    \end{subfigure}
    
    \caption{Selection Ratio}
    \label{fig:side_by_side2}
\end{figure}


From these plots, a few clear patterns stand out:\\

\noindent Increasing the \textbf{number of experts} strengthens the \textbf{skill component}, as analytical participants can better exploit differences in team quality, and mean winnings for the best team are higher in the case of 4 experts. In general, more choices allow participants to think strategically, increasing the role of \textbf{skill} in the game.  \\

\noindent Intuitively, \textbf{higher correlations} should make it more difficult to guess the best team and thus reduce the \textbf{skill component} of the game. However, in \textbf{unequal mean} cases, if the correlation is very high, the teams can become clearly differentiated, making it \textbf{paradoxically easier} to pick. For example, if two teams share ten players but differ in \textbf{Rohit Sharma} (top-order batsman) versus \textbf{Rinku Singh} (finisher), participants would pick the team with Rohit Sharma in it. Thus, \textbf{correlation} does play a role, but its effect depends on the differences between key players in the teams.

Taken together, these results highlight that both the \textbf{number of experts} and the \textbf{degree of correlation} are structural features that directly shape the balance between \textbf{skill} and \textbf{luck} in such games.

\subsection{Shift in Game Dynamics with \textbf{Impact Player}}
\label{sec:theo_impact}

The introduction of the Impact Player provides a significant performance boost to the user's team, making the decision of whom to select an important and strategic choice. By adding this layer of consideration, the feature encourages more careful planning and foresight, thereby enhancing the skill-based component of the game. Players must analyze team composition, potential contributions, and the pool of available impact players, which collectively introduces a higher degree of user-side optimization and strategic depth.

Keeping consistency with our earlier setups, we assume that there is only a set of 4 impact players, from which people choose. This is a reasonable assumption in most cricket matches as usually, the best cricket players are included in the expert team, and from the remaining pool, there is usually a limited set of players who have the potential to outshine other players. The impact players points are jointly modelled as 
\[
(I_1,I_2,I_3,I_4)\sim\text{LogNormal}(\boldsymbol{\mu_I}, \boldsymbol{\Sigma_I})
\] For simplifying the setup, $\Sigma_I$ is assumed to have the equicorrelation structure, with the mild correlation of $\rho_I=0.3$. Also, the impact player points vector \textbf{I}  is assumed to be independent of the team points vector \textbf{P}. 

We model the proportion of contribution of the lowest-performing player in the \(i^{\text{th}}\) team by the random variable \(S_i\). 
To characterize its distribution, let
\[
X = (X_1, X_2, \ldots, X_{11}) \sim \text{Dirichlet}(\boldsymbol{\alpha}),
\]
represent the vector of normalized contribution proportions within a team. 
We define
\[
S = \min_{1 \leq k \leq 11} X_k,
\]
as the minimum coordinate of the Dirichlet random vector, corresponding to the contribution proportion of the lowest-performing player.  
We refer to the distribution of \(S\) as the \emph{Dirichlet–Minimum distribution}, denoted by
\[
S \sim \text{DirichletMin}(\boldsymbol{\alpha}).
\]
Accordingly, the team-specific proportions are modeled as independent and identically distributed realizations from this distribution:
\[
S_i \stackrel{\text{i.i.d.}}{\sim} \text{DirichletMin}(\boldsymbol{\alpha}), 
\qquad i = 1, 2, \ldots, n.
\]
For our simulations, the Dirichlet concentration parameter has been set to 
\[
\boldsymbol{\alpha} = 10\,\mathbf{1},
\]
where \(\mathbf{1}\) denotes a vector of ones of dimension equal to the number of players.

Let $B_{ij}$ be the boosted score of the $i^{th}$ expert team with the $j^{th}$ impact player.
\[
B_{ij} =  P_i(1-S_i)+max(S_i\cdot P_i, I_j)
\]
As before, we need to find the win probability of each team-impact player combination. Consistent with our earlier notation, we assume $\pi_{ij}$ is the probability that the combination of the $i^{th}$ expert team and the $j^{th}$ impact player achieves the maximum score among 
all team-impact player combinations, in essence meaning that the combination wins the contest
\[
\pi_{ij} = \mathbb{P}(B_{ij} \ge B_{kl} \text{ } \forall (k,l) \neq (i,j))
\]

As before, these probabilities are difficult to compute analytically and thus we resort to simulation based approaches, where we draw repeated samples from the team points and impact player points distribution to estimated each team-impact player combination's probability of winning the contest.
\\
In order to observe the changes in the game dynamics brought in by the introduction of the choice of impact player, we use three different configurations of impact players.

\begin{table}[h]
\centering
\begin{tabularx}{\textwidth}{@{} l c c X @{}}
\toprule
\textbf{Name} & \textbf{Mean Points} & \textbf{Std. Deviations} & \textbf{Interpretation} \\
\midrule
IID & $[45,45,45,45]$ & $[10,10,10,10]$ & All impact players are statistically identical in both mean and variance (in essence, making the effect of impact players uniform). \\
\addlinespace
Different\_mean & $[35,40,45,50]$ & $[10,10,10,10]$ & Impact Players differ in average performance (mean) but maintain equal variability (standard deviation). \\
\addlinespace
Different\_mean\_and\_std & $[35,40,45,50]$ & $[5,10,15,20]$ & An asymmetric setting where impact players vary both in average strength and consistency, reflecting diverse abilities. \\
\bottomrule
\end{tabularx}
\caption{Impact Player Configurations}
\label{tab:impactconfigurations}
\end{table}

Now, in order to assess whether impact players truly add an element of skill to the game, we need to examine whether asymmetries in the skills of the impact players introduce an additional strategic dimension—making the selection of the correct impact player a crucial decision. For this purpose, we plot the mean deviations of each team–impact player combination across all the team point configurations defined earlier in the paper.

We compute the mean deviation for a particular expert–impact player combination within a team point configuration by taking the average winnings of the expert in that configuration and subtracting it from the mean winnings of the corresponding expert–impact player combination.
From the below plots, we can observe that the deviations are quite high in the Different Mean and Different Mean and Std configurations as compared to the IID configuration. This indicates that a greater degree of variability is injected in the game by inclusion of the Impact Player, thus allowing more skillful players to gain an advantage in the contest structure.
\begin{figure}[H]
    \centering
    \begin{minipage}{0.48\textwidth}
        \centering
        \includegraphics[width=\textwidth]{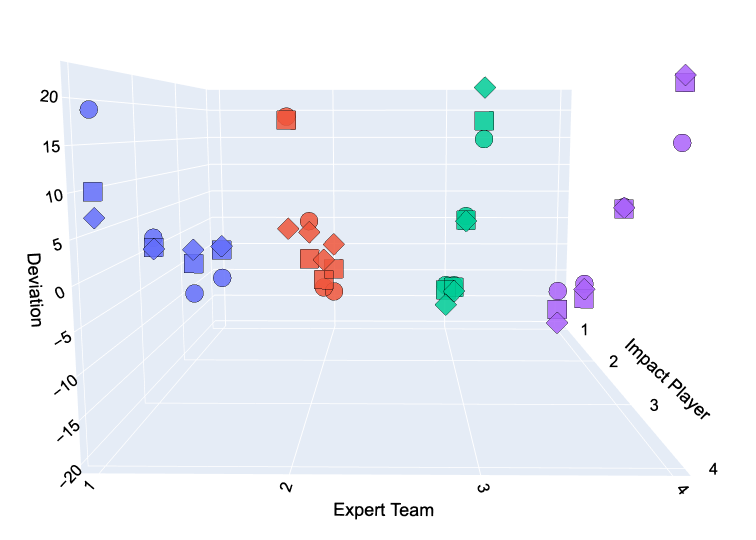}
        \caption{Different Mean configuration.}
        \label{fig:different-mean}
    \end{minipage}%
    \hfill
    \begin{minipage}{0.48\textwidth}
        \centering
        \includegraphics[width=\textwidth]{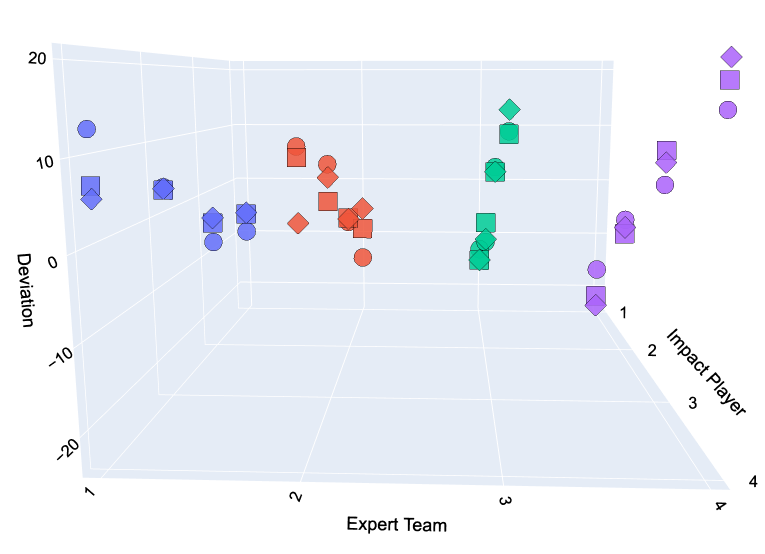}
        \caption{Different Mean and Std configuration.}
        \label{fig:different-mean-std}
    \end{minipage}
\end{figure}

\begin{figure}[H]
    \centering
    \includegraphics[width=0.6\textwidth]{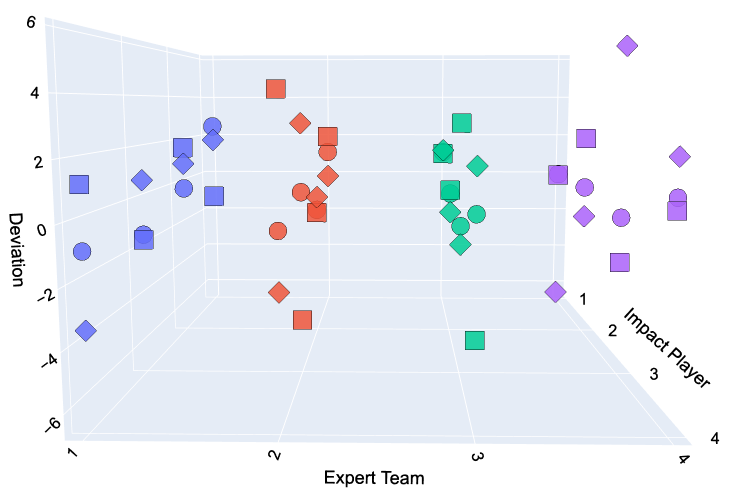}
    \caption{IID configuration.}
    \label{fig:IID}
\end{figure}
\begin{figure}[h!]
    \centering
    \includegraphics[width=1\textwidth]{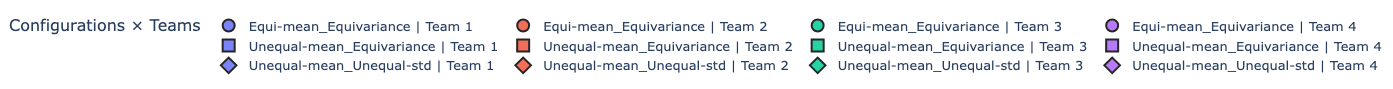}
    \caption{Legend}
    \label{fig:legend}
\end{figure}

To quantify the increase in variability, we calculate the \( F \)-statistic using the following formula:
\[
F = \frac{\text{Total Variance for the particular impact player configuration}}{\text{Total Variance for the IID configuration}}
\]

Higher values of the \( F \)-statistic imply that asymmetries in the impact player’s skill meaningfully influence outcomes. This, in turn, suggests that the inclusion of impact players enhances the role of skill in the game.

The computed \( F \)-statistics are as follows:
\begin{itemize}
    \item \textbf{Different\_mean}: 28.18735
    \item \textbf{Different\_mean\_and\_std}: 22.63239
\end{itemize}

Both values are considerably greater than 1, indicating that the inclusion of Impact Players introduces a substantial increase in the variability of mean winnings across simulated contests. 
This heightened dispersion implies that participant outcomes become more sensitive to underlying decision quality, as the selection or utilisation of the Impact Player now contributes materially to overall performance. 
In other words, the introduction of this tactical feature amplifies the consequences of strategic choices, thereby strengthening the manifestation of skill within the game and reducing the relative influence of random fluctuations.

\section{Implementation with Real Life Data}
\label{sec:real}

Up to this point, we have examined the game format under various simulation configurations and gathered substantial evidence that the \textbf{Expert of Experts (EOE)} contest contains a meaningful component of skill. In this section, we extend the analysis to real-life settings and explore how these assertions hold in practice. We also outline a suitable protocol for implementing the game.  

In the proposed real-life implementation of the EOE contest, each participant is required to select one team from a pool consisting of 2--4 expert-designed teams. Entry into the contest is conditional on the payment of a participation fee of Rs.~25 per user. The aggregate entry fees collected from all users constitute the initial prize pool, from which the platform charges a fixed service fee of 20\% as its operational cut. The remaining 80\% of the pool forms the distributable prize fund. Consistent with the \textbf{Shared Prize} format outlined earlier, this fund is distributed equally among all participants who chose the expert team that eventually secures the highest score in the underlying real-world match. Participants whose selected team does not achieve the highest score are not entitled to any share of the prize pool. This structure ensures that the game remains simple to understand while also providing clear incentives for participants to evaluate expert teams carefully, as their returns depend entirely on selecting the top-performing option.  

In this setup, a \textbf{Player} refers to the actual cricket players competing in the real-world match, whereas a \textbf{User} refers to a participant in the EOE game. The primary objective of this analysis is to evaluate whether participants who employ a structured, rule-based strategy achieve outcomes that are consistently superior to random selection. Demonstrating such a systematic advantage would provide empirical evidence that the \textbf{Expert of Experts (EOE)} game rewards informed decision-making rather than pure chance.  

Our analysis draws on two primary categories of data. \textbf{Match Scorecards (IPL 2024)} were retrieved via the Cricbuzz Rapid API, providing detailed information on fixtures, team outcomes, and granular player-level performance (batting, bowling, and fielding contributions). \textbf{Player Career Statistics} were compiled from ESPN Cricinfo, including historical records across formats, aggregate career averages, and totals. 

The collected data were organized into structured datasets suitable for computational analysis. \textbf{Tournament Data} includes the 71 completed matches of IPL 2024, preprocessed into a single \texttt{DataFrame}, capturing the progression of the tournament in a machine-readable format. \textbf{Player Data} contains all 263 registered players, with their career-level statistics collated in a separate \texttt{DataFrame} to enable integration of historical performance with tournament-specific outcomes.  

\begin{figure}[H]
    \centering
    \includegraphics[width=0.9\linewidth]{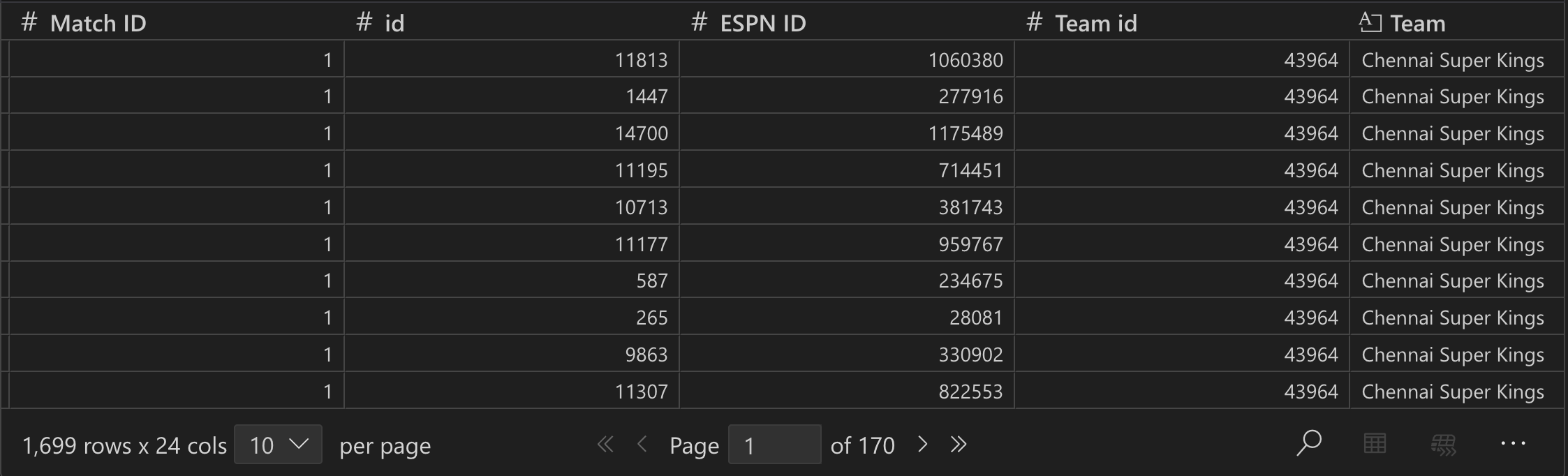}
    \caption{Snapshot of the Scorecard \texttt{DataFrame}. Each row corresponds to a player-match instance with associated performance metrics.}
    \label{fig:scorecard}
\end{figure}

\begin{figure}[H]
    \centering
    \includegraphics[width=0.9\linewidth]{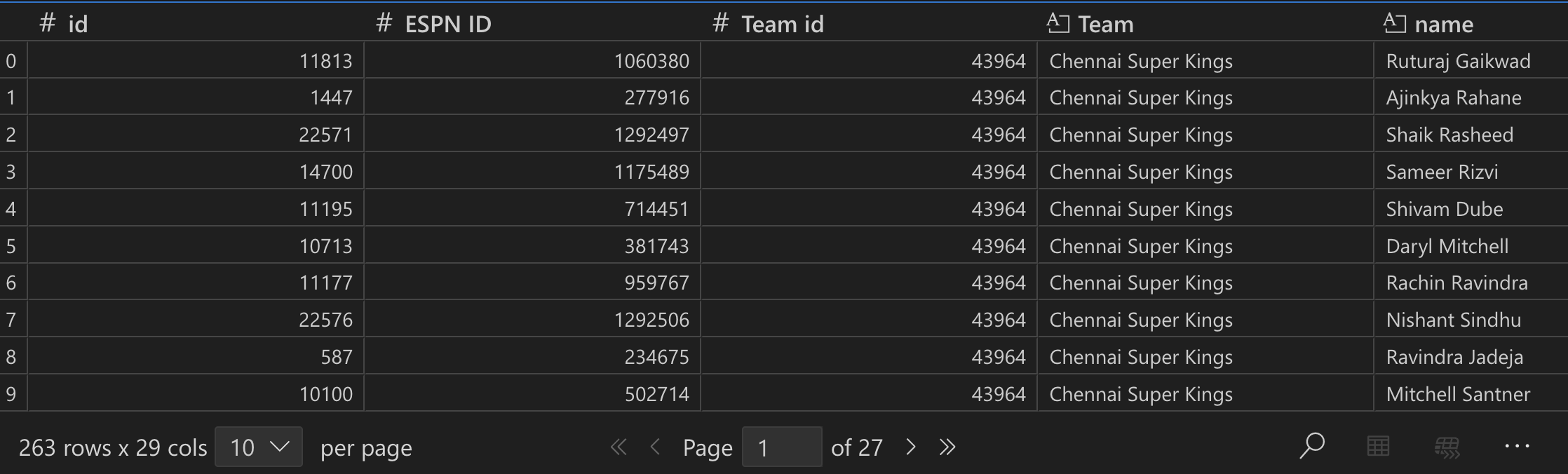}
    \caption{Snapshot of the Squads \texttt{DataFrame}. Each row represents a registered player along with demographic details and career-level statistics.}
    \label{fig:squads}
\end{figure}

\subsection{Performance Metrics, Expert Strategies, and User Behaviour Modeling}

To quantitatively assess individual player performance, three key metrics were defined to capture different dimensions of ability and consistency. The first is \textbf{Career Points}, which represent the long-run performance level of a player. This metric is computed as the mean of the historical \textit{My11Circle} points obtained by the player, based on the scoring rules applicable as of November 2024. In cases where one or more sub-metrics required for this calculation were unavailable, the corresponding entries were uniformly disregarded across all players to maintain comparability. 

The second measure, \textbf{Form}, reflects a player’s recent performance trend. For a given window of $n$ matches, it is calculated as the average of the \textit{My11Circle} points earned by the player across their most recent $n$ appearances. This metric highlights short-term fluctuations in performance that may not be fully captured by long-run averages. Finally, \textbf{Tournament Points} quantify a player’s performance within a specific competition. It is obtained as the average of the points accumulated by the player in all matches of the relevant tournament. Together, these three metrics provide a comprehensive evaluation framework, balancing long-term stability, short-term form, and tournament-specific performance to represent the multidimensional nature of player skill.

Building on these definitions, the dynamics of the \textbf{Expert of Experts (EOE)} contest were replicated through a simulation framework based on the final ten completed matches of the IPL 2024 season. Within this framework, a set of \emph{expert teams} was generated using automated \emph{strategy bots} that operationalized different selection heuristics grounded in the performance metrics defined above. Two categories of user types were then modeled: \textbf{analytical users}, who employed structured decision rules based on these metrics, and \textbf{random users}, who selected teams uniformly at random. This setup enabled direct comparison between strategic and non-strategic participation under controlled contest conditions.

Each pair of expert teams was designed to share at least $n$ common players ($n \in \{5,6,7,8,9\}$), ensuring that observed differences across teams emerged primarily from differences in strategic emphasis rather than arbitrary composition. The canonical core of each team was first identified by solving a linear programming problem to satisfy MEC (Minimum Expert Consensus) constraints while maximizing \textbf{Career Points}. The remaining $(11-n)$ slots were then filled according to the particular strategy rule, with Captain and Vice-Captain roles assigned following the same selection logic.

Four primary expert strategies were implemented to capture distinct evaluation philosophies. Under the \textbf{Career Points Strategy}, players were ranked according to their historical \textbf{Career Points}, with the top $(11-n)$ selected and the two highest-scoring players designated as Captain and Vice-Captain. The \textbf{MA5 Strategy} ranked players based on their average \textbf{Form} over the most recent five matches, following the same leadership assignment. The \textbf{Tournament Points Strategy} relied on the player’s mean performance within the IPL 2024 tournament, again assigning leadership roles to the top two by that metric. Finally, the \textbf{Mean–Variance Optimization (MeanVarOptim)} approach, inspired by the integer optimisation framework proposed by \cite{b2}, evaluated player performance through a risk-sensitive mean statistic:
\[
\textbf{MeanVar} = \text{Form (over last 3 matches)} - \lambda \cdot \text{SD(points over last 3 matches)},
\]
where $\lambda = 0.5$ represents the penalty for volatility in recent performance. Players with the highest \textbf{MeanVar} scores were selected to fill the remaining positions, with the top two assigned as Captain and Vice-Captain. This multi-strategy structure ensured that all expert teams shared a common backbone while still reflecting distinct interpretations of player value and risk preference.

In contrast, user teams were constructed to represent varying degrees of strategic engagement. \textbf{Skillful users} followed deterministic selection rules based on different weighted combinations of \textbf{Form} and \textbf{Career Points}, while \textbf{random users} selected among expert teams uniformly at random. Three specific deterministic rules were implemented to model different cognitive weighting schemes. The \textbf{One-Third Form} rule emphasized long-term consistency, defined as
\[
\textbf{OneThirdForm} = \tfrac{1}{3} \cdot \textbf{Form} + \tfrac{2}{3} \cdot \textbf{Career Points},
\]
whereas the \textbf{Two-Third Form} rule prioritized recent performance, given by
\[
\textbf{TwoThirdForm} = \tfrac{2}{3} \cdot \textbf{Form} + \tfrac{1}{3} \cdot \textbf{Career Points}.
\]
Finally, the \textbf{Strict Form} rule was purely form-driven, relying exclusively on recent match performance:
\[
\textbf{StrictForm} = \textbf{Form}.
\]
These formulations collectively allowed the simulation to capture a realistic spectrum of user behaviour, ranging from fully random to analytically optimized decision-making, thereby facilitating a systematic evaluation of skill expression within the limited-selection fantasy contest environment.  

To incorporate an element of real-world imperfection and bounded rationality in decision-making, we introduce a stochastic perturbation: with probability $p = 0.05$, the user disregards the prescribed deterministic rule and instead selects a team at random. This randomization step prevents the simulated users from behaving as perfectly rational agents and better reflects observed patterns in practical fantasy gameplay.

\subsection{Evaluation Metrics}

We vary the \textbf{number of expert teams} ($k \in \{2,3,4\}$) and the \textbf{number of common players} ($n \in \{5,6,7,8,9\}$). Each configuration is run 10,000 times with 1,000 skill-based users per strategy and 16,000 random users.  

To systematically assess the relative performance of strategy-driven users versus random selectors, we define a set of evaluation metrics. These metrics are designed to capture both match-level and tournament-level differences, as well as to provide evidence of whether skillful decision-making translates into measurable advantages. Specifically:  

\begin{itemize}  
    \item \textbf{Average Gain over Random (Matchwise):} This metric computes, on a per-match basis, the mean difference in accrued points between a given skill-based strategy and the random user baseline. A positive value indicates that, on average, the strategy outperforms random selection in individual matches.  

    \item \textbf{Average Gain over Random (Tournament):} Extending the above, this aggregates the gains across all matches in the tournament. It reflects the cumulative advantage of a strategy relative to random play over the entire season, thereby offering a holistic measure of long-run performance.    
\end{itemize}  

Together, these metrics allow us to evaluate not only point-based gains but also behavioral indicators of skill, thereby providing a comprehensive framework for comparing structured strategies against random selection.

\subsection{Observations}

     Across the simulated scenarios, the overall presence of skill in this particular setup appears to be \textbf{limited}. In many configurations, the average gain achieved by users employing skill-based strategies is negative relative to random selection, indicating that, on average, even carefully designed strategies often underperform compared to a purely random approach.
     
     Increasing the number of common players shared among expert teams tends to improve the average gain over random selection up to a certain point, after which the improvement starts to decline. This pattern suggests that when the pool of choices becomes somewhat simplified due to greater overlap in team compositions, the signal of skill or strategy becomes clearer, allowing informed decisions to perform better than random ones. However, as the overlap keeps increasing and the teams become too similar, the available information becomes less distinctive, making it harder to achieve performance that is meaningfully better than random selection. 
     
     Furthermore, as the number of expert teams available to users grows, the average gain over random selection generally increases. This reflects the notion that a wider array of choices provides participants with more opportunities to exercise skill and differentiate between teams based on the available information.

     Another observation which can be made here is that the Two-Third Form and the Strict Form user strategies are very similar in their mean winnings but differ significantly as compared to the One-Third Form strategy. So, our analysis also needs to be tuned according to the user strategies. 
     
     Taken together, these observations reveal a nuanced relationship between the structure of available choices and the expression of skill. Increasing the number of experts expands the opportunity for participants to demonstrate strategic ability, but the degree of similarity among expert teams strongly influences how visible that skill becomes. When multiple experts share overlapping player selections or adopt similar reasoning, the choice space becomes more constrained and the underlying signals more coherent. In such settings, skillful strategies can more effectively exploit these clearer signals, leading to noticeable performance gains. However, as overlap continues to grow and expert compositions become nearly identical, the diversity of available information diminishes, making it increasingly difficult for any strategy to outperform random selection.

\begin{figure}[H]
    \centering
    \begin{subfigure}[b]{0.45\textwidth}
        \centering
        \includegraphics[width=\textwidth]{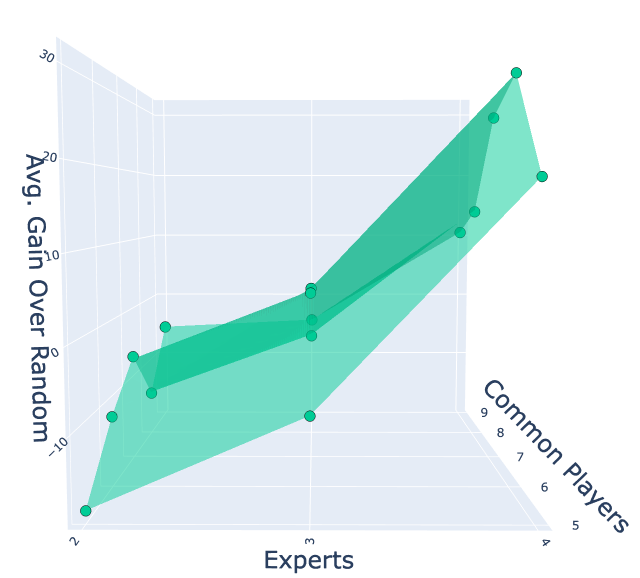}
        \caption{Across experts}
    \end{subfigure}
    \hfill
    \begin{subfigure}[b]{0.45\textwidth}
        \centering
        \includegraphics[width=\textwidth]{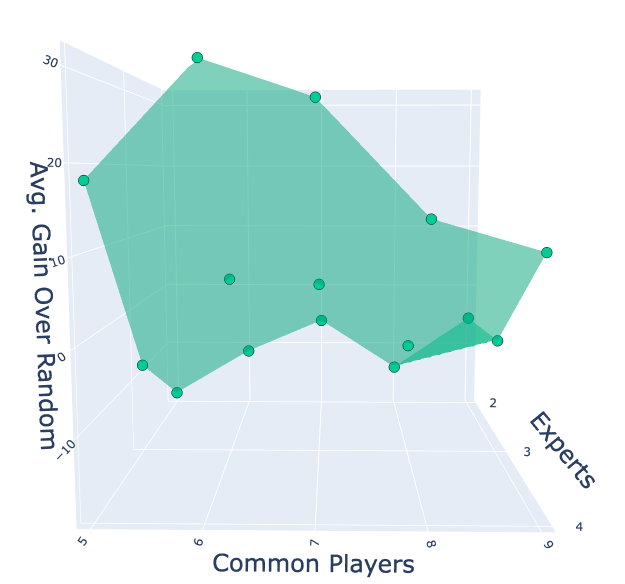}
        \caption{Across common players}
    \end{subfigure}
    \caption{Empirical distribution under the One-Third Form.}
    \label{fig:empirical_one_third1}
\end{figure}

\begin{figure}[H]
    \centering
    \begin{subfigure}[b]{0.45\textwidth}
        \centering
        \includegraphics[width=\textwidth]{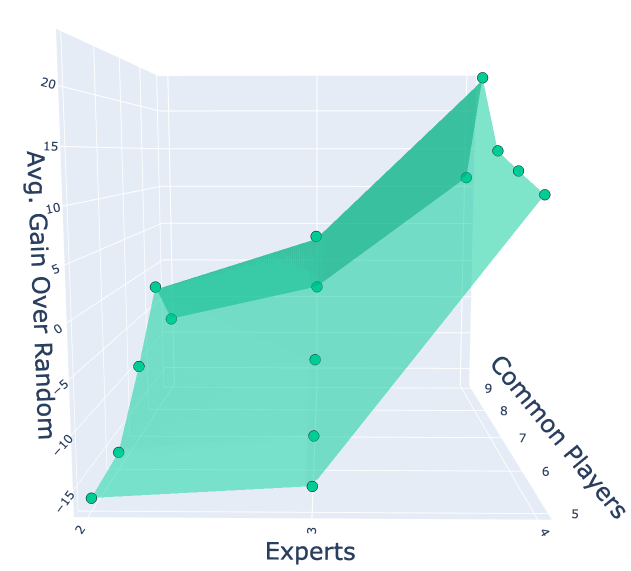}
        \caption{Across experts}
    \end{subfigure}
    \hfill
    \begin{subfigure}[b]{0.45\textwidth}
        \centering
        \includegraphics[width=\textwidth]{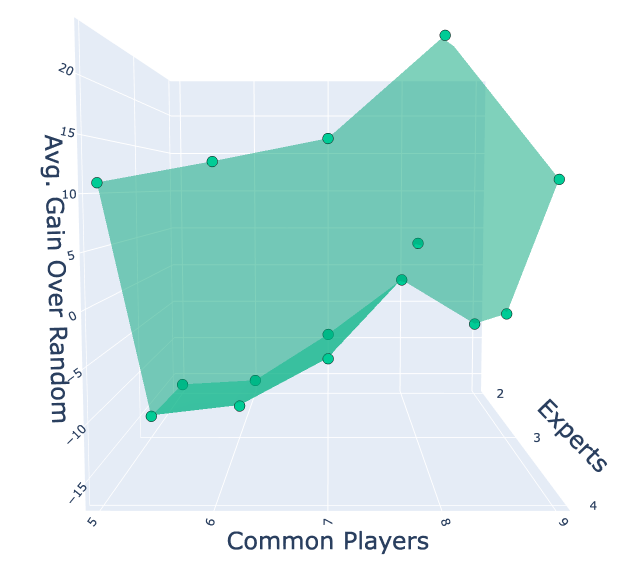}
        \caption{Across common players}
    \end{subfigure}
    \caption{Empirical distribution under the Two-Third Form.}
    \label{fig:empirical_two_third1}
\end{figure}

\begin{figure}[H]
    \centering
    \begin{subfigure}[b]{0.45\textwidth}
        \centering
        \includegraphics[width=\textwidth]{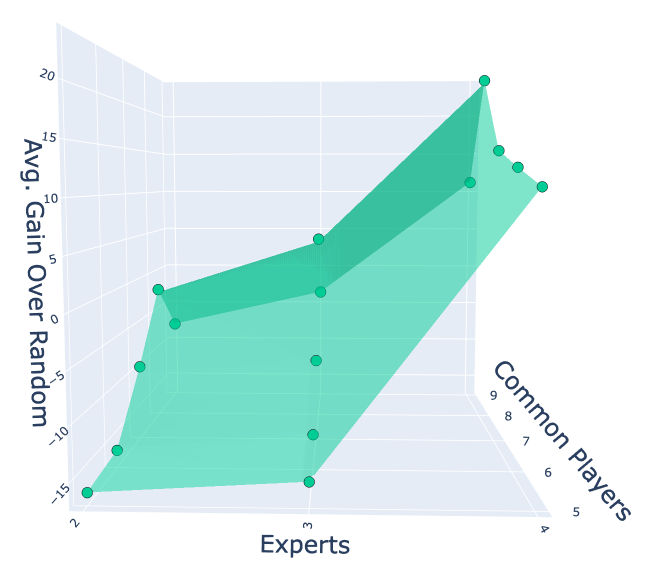}
        \caption{Across experts}
    \end{subfigure}
    \hfill
    \begin{subfigure}[b]{0.45\textwidth}
        \centering
        \includegraphics[width=\textwidth]{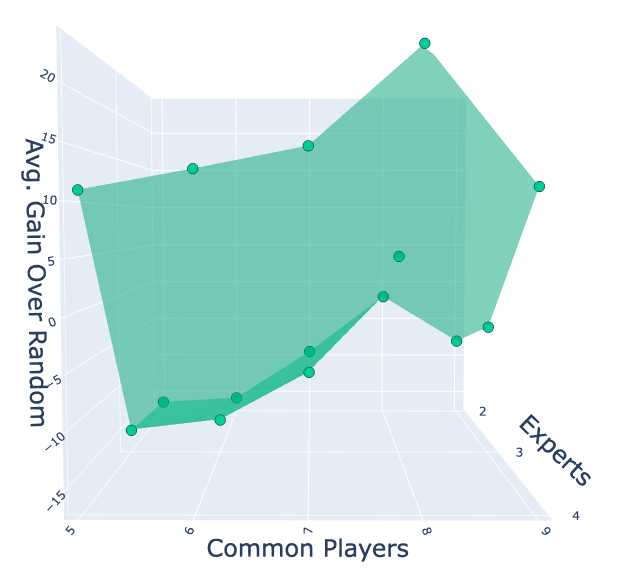}
        \caption{Across common players}
    \end{subfigure}
    \caption{Empirical distribution under the Strict Form.}
    \label{fig:empirical_strict1}
\end{figure}

\subsection{Effect of Impact Player}
\label{sec:real_impact}
The incorporation of an \textbf{impact player} introduces an additional lever of control for users, allowing them to potentially influence team performance more directly. This was heoretically discussed in section \ref{sec:theo_impact}. The procedure for integrating impact players with real data is now attempted as follows:  

First, expert teams are generated using the previously described methodology. Subsequently, each expert team is systematically augmented by considering, one at a time, each player from the remaining pool in that match as a potential impact player. This results in a set of modified teams, each incorporating a different candidate for the impact player role.  

Once these augmented teams are constructed, the predefined user strategies are applied to evaluate performance. The evaluation process differs from the standard setup in only one specific respect: if the player in a team with the lowest score performs worse than the designated impact player, the contribution of that lowest-performing player is replaced with the contribution of the impact player.  

All other aspects of the analysis, including metrics, simulation procedures, and comparison with random or other skillful strategies, are conducted identically to the non-impact-player scenario. This ensures that the additional degree of freedom provided by impact players can be assessed in a consistent and controlled manner, allowing for a rigorous evaluation of their influence on the effectiveness of user strategies.

\subsection*{Illustrative plots and observations}
First, the inclusion of an impact player appears to significantly enhance the performance of the skill-based (or strategic) team selection mechanisms relative to the random strategies. This improvement is evident from the shift of the gain distributions toward positive values, indicating that the presence of an impact player amplifies the influence of skill and informed decision-making in determining outcomes. In other words, when the impact player mechanism is introduced, the distinction between skilled and random strategies becomes more pronounced, with skillful strategies demonstrating a clear advantage.\\
 As in the Basic case, if the total no. of experts increase, the average gain tends to rise consistently, showing that a broader pool of expert of opinions allow for strategic users to gain predictive advantage over the random users.\\

However in contrast with the Basic case, if the no. of common players is increased, the average gain first decreases upto a certain point after which it starts improving.This seems contradictory with our earlier ideas. This abberation can be due to the fact that when skillful strategies win, they usually do that by a wide margin. So, the averaging might lead to confounding of the actual variable. Some further comments are made on this in the Regression Analysis section.

\begin{figure}[H]
    \centering
    \begin{subfigure}[b]{0.45\textwidth}
        \centering
        \includegraphics[width=\textwidth]{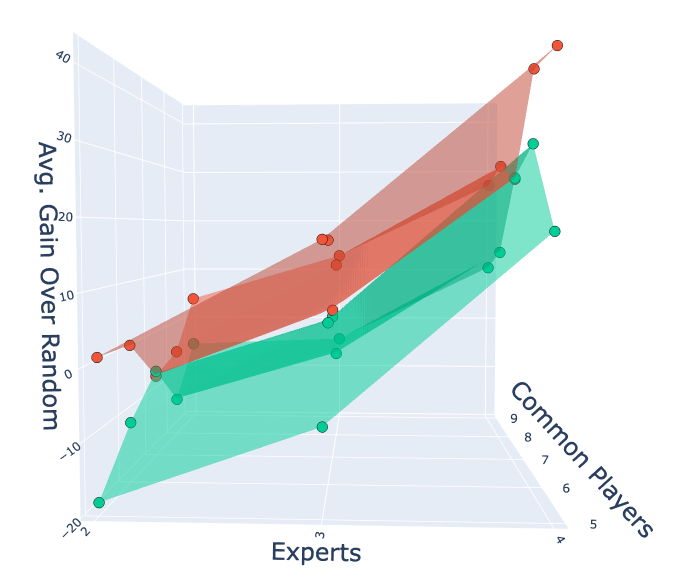}
        \caption{Across experts}
    \end{subfigure}
    \hfill
    \begin{subfigure}[b]{0.45\textwidth}
        \centering
        \includegraphics[width=\textwidth]{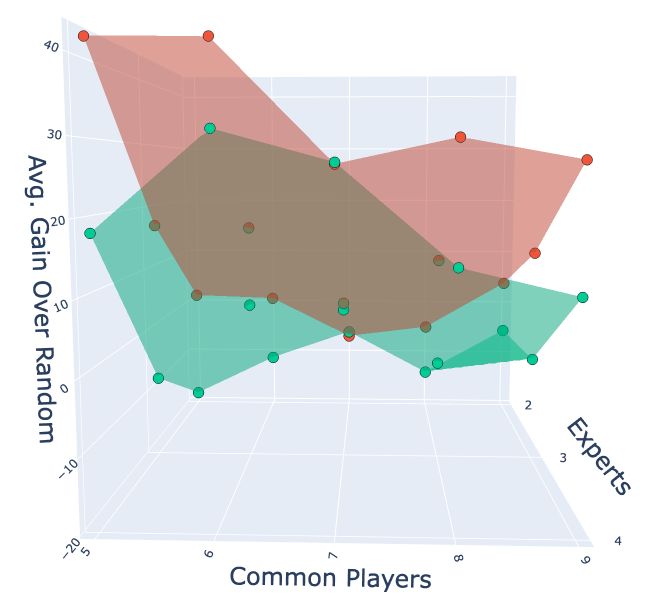}
        \caption{Across common players}
    \end{subfigure}
    \caption{Empirical distribution under the One-Third Form.}
    \label{fig:empirical_one_third2}
\end{figure}

\begin{figure}[H]
    \centering
    \begin{subfigure}[b]{0.45\textwidth}
        \centering
        \includegraphics[width=\textwidth]{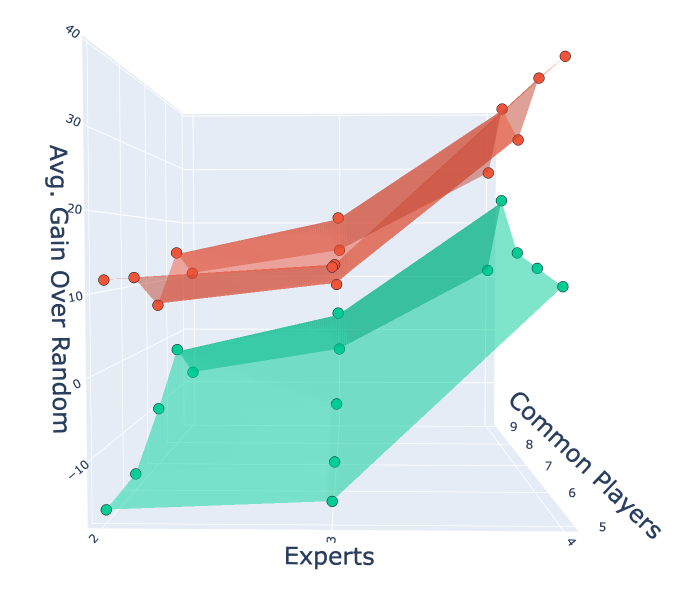}
        \caption{Across experts}
    \end{subfigure}
    \hfill
    \begin{subfigure}[b]{0.45\textwidth}
        \centering
        \includegraphics[width=\textwidth]{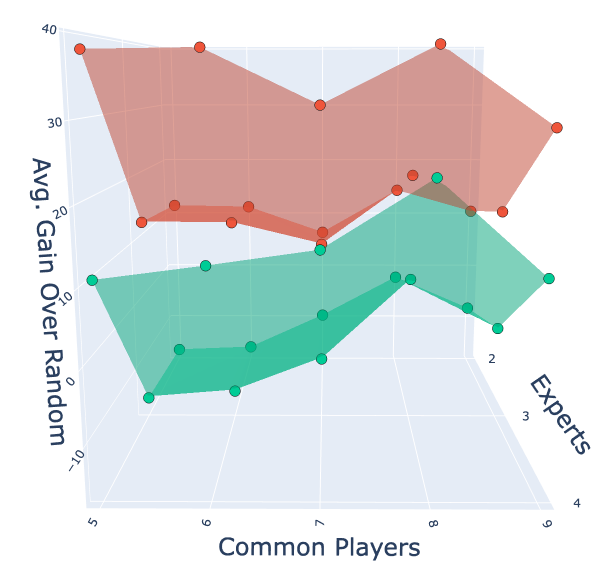}
        \caption{Across common players}
    \end{subfigure}
    \caption{Empirical distribution under the Two-Third Form.}
    \label{fig:empirical_two_third2}
\end{figure}

\begin{figure}[H]
    \centering
    \begin{subfigure}[b]{0.45\textwidth}
        \centering
        \includegraphics[width=\textwidth]{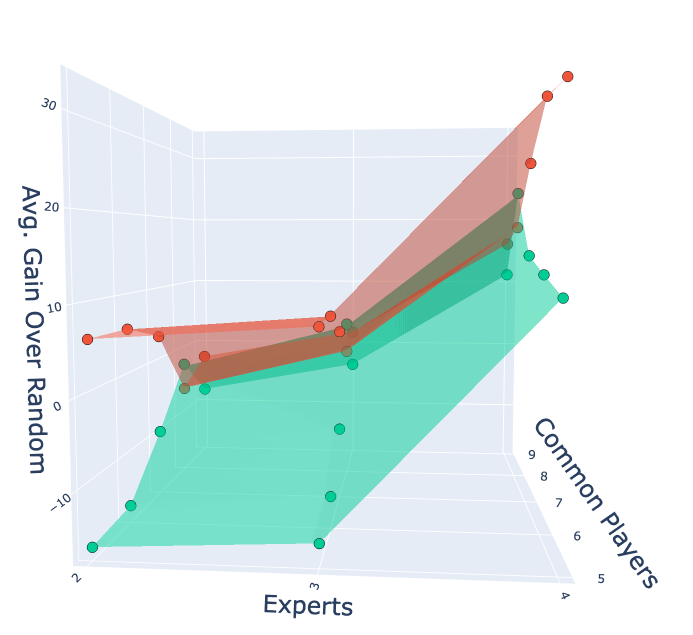}
        \caption{Across experts}
    \end{subfigure}
    \hfill
    \begin{subfigure}[b]{0.45\textwidth}
        \centering
        \includegraphics[width=\textwidth]{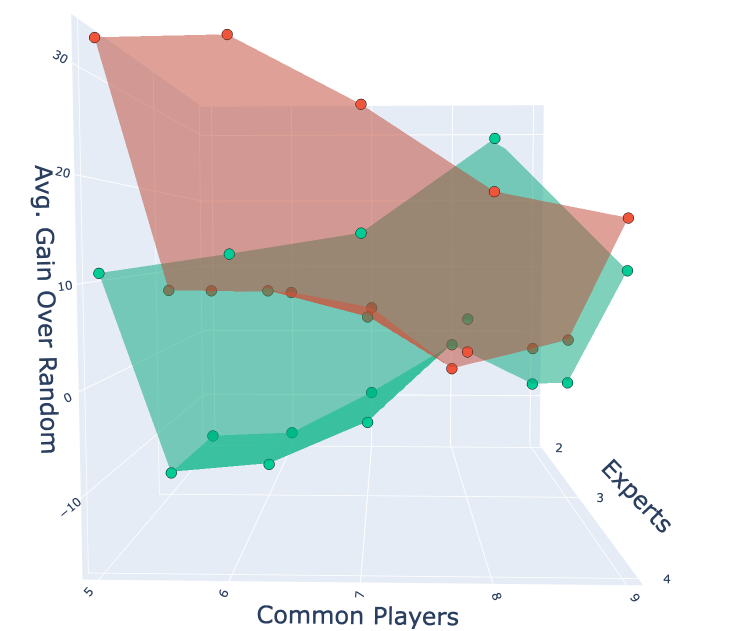}
        \caption{Across common players}
    \end{subfigure}
    \caption{Empirical distribution under the Strict Form.}
    \label{fig:empirical_strict2}
\end{figure}

\section{Cause and Effect Analysis}
\label{sec:regression}
Till now, we have made various observations which indicate that there is significant presence of skill in the game. So, we fit various regression models, to quantitatively model the dependence of skill on various factors to test whether these factors are indeed significant or not.

\subsection{Model Specification}

To investigate the drivers of performance in the \textit{Expert-of-Experts} fantasy game, we construct regression models that relate average strategic gains to design and skill-related covariates.  
The model specification is given by:
\[
G_i = \beta_0 + \beta_1 E + \beta_2 C + \beta_3 V_\mu + \beta_4 \overline{\Sigma}^2 + \varepsilon_i,
\]
where:
\begin{itemize}
    \item $G_i$ denotes the average gain for a given configuration and strategy,
    \item $E$ is the number of expert teams,
    \item $C$ (or $C_p$) represents the number of (actual or prescribed) common players,
    \item $V_\mu$ is the between-team variability of points,
    \item $\overline{\Sigma}^2$ denotes within-team variability.
\end{itemize}

The analysis is carried out for two gameplay configurations — \textbf{Basic} and \textbf{Impact Player (IP)} — and for two strategies: \textit{Strict on Form} and \textit{One-Third Form}.  
All models are estimated using Ordinary Least Squares (OLS).

\subsection{Descriptive Findings}

Descriptive statistics indicate that the presence of the Impact Player substitution amplifies skill-driven outcomes.  
The mean gain across strategies increases consistently with the number of expert teams, suggesting that a richer expert pool enhances opportunities for differentiation.

Correlation analysis further reveals:
\begin{itemize}
    \item $V_\mu$ and $\overline{\Sigma}^2$ both influence gains.
    \item Within-team and between-team variances affect the “one-third” strategy in opposite directions, reflecting distinct dynamics in that decision rule.
\end{itemize}
For a more detailed analysis, we cross tabulate the average gain across vaying no. of experts and prescribed no. common players.

With a higher number of teams, random choices are more strongly dominated by strategic choices. In fact, for few experts, the random strategy actually has a higher average (strategic gain is negative).

 Cross-tabulation  of the gain across prescribed no. of common players show that with an increase of common players, the game becomes manageable for strategic players and hence gain increases. But once commonality becomes too high, the analytical flavour of the game is gradually lost and the strategic players’ advantage diminish. Again, this proves the presence of skill.

\subsection{Regression Results}
\textbf{Linear Model results} 

\noindent The results consistently show a significantly "positive" relationship of Average gain with the no. of experts.

\noindent The no. of actual common players (or prescribed common players), as expected has lower impact in the linear form, as we anticipated an increasing then decreasing decreasing relation from the cross tabulations.

\noindent Variability between teams $V_{\mu}$ has a generally positive impact on gain when significant, while the within-group variation  $\overline{\Sigma}^2$  is generally less impactful in the presence of other covariates.
The tables present regression coefficients and significance levels for the various model specifications.
\begin{table}[H]
\centering
\label{tab:reg_quad1}
\begin{tabular}{lcccc}
\toprule
\textbf{Dependent Variable: StrictOnForm Gain} & \textbf{Coefficients} & \textbf{Sig.} & \textbf{Coefficients} & \textbf{Sig.} \\
\midrule
(Constant) & -195.341 & 0.000 & -236.196 & 0.000 \\
Prescribed Common Players ($C_p$) & 9.424 & 0.000 & --- & --- \\
Actual Common Players ($C$) & --- & --- & 12.480 & 0.021 \\
Experts ($E$) & 20.737 & 0.000 & 21.217 & 0.000 \\
$V_\mu$ & 0.030 & 0.000 & 0.028 & 0.000 \\
$\overline{\Sigma}^2$ & 0.000479 & 0.000 & 0.001 & 0.000 \\
$R^2$ & 0.355 &  & 0.313 &  \\
$F$-statistic & 19.965 &  & 16.55 &  \\
\bottomrule
\end{tabular}
\end{table}
\vspace{1em}
\begin{table}[h!]
\centering
\label{tab:reg_quad2}
\begin{tabular}{lcccc}
\toprule
\textbf{Dependent Variable: OneThirdForm Gain} & \textbf{Coefficients} & \textbf{Sig.} & \textbf{Coefficients} & \textbf{Sig.} \\
\midrule
(Constant) & -124.380 & 0.031 & -34.633 & 0.218 \\
Experts ($E$) & 17.839 & 0.000 & 15.558 & 0.000 \\
Prescribed Common Players ($C_p$) &--  &-- & 0.982 & 0.710 \\
Actual Common Players ($C$) & 10.014 & 0.075 & -- & -- \\
$V_\mu$ & 0.013 & 0.080 & 0.005 & 0.429 \\
$\overline{\Sigma}^2$ & 0.000 & 0.005 & -0.000299 & 0.015 \\
$R^2$ & 0.141 &  & 0.159 &  \\
$F$-statistic & 5.943 &  & 6.840 &  \\
\bottomrule
\end{tabular}
\end{table}

\vspace{1em}
\begin{table}[h!]
\centering
\label{tab:reg_quad3}
\begin{tabular}{lcccc}
\toprule
\textbf{Dependent Variable: IP\_StrictOnForm Gain} & \textbf{Coefficients} & \textbf{Sig.} & \textbf{Coefficients} & \textbf{Sig.} \\
\midrule
(Constant) & -225.042 & 0.000 & -115.801 & 0.000 \\
Experts ($E$) & 21.727 & 0.000 & 19.109 & 0.000 \\
Prescribed Common Players ($C_p$) &--  &-- & 3.727 & 0.177 \\
Actual Common Players ($C$) & 14.423 & 0.014 & -- & -- \\
$V_\mu$ & 0.038 & 0.000 & 0.031 & 0.080 \\
$\overline{\Sigma}^2$ & 0.000 & 0.417 & 0.0 & 0.234 \\
$R^2$ & 0.236 &  &  &  \\
$F$-statistic & 11.182 &  &  &  \\
\bottomrule
\end{tabular}
\end{table}

\vspace{1em}
\begin{table}[h!]
\centering
\label{tab:reg_quad4}
\begin{tabular}{lcccc}
\toprule
\textbf{Dependent Variable: IP\_OneThirdForm Gain} & \textbf{Coefficients} & \textbf{Sig.} & \textbf{Coefficients} & \textbf{Sig.} \\
\midrule
(Constant) & -26.035 & 0.344 & -93.388 & 0.101 \\
Experts ($E$) & 17.056 & 0.000 & 19.001 & 0.000 \\
Prescribed Common Players ($C_p$) &-3.025  &0.242 & -- & -- \\
Actual Common Players ($C$) & -- & -- & 4.196 & 0.449 \\
$V_\mu$ & -0.005 & 0.468 & 0.004 & 0.621 \\
$\overline{\Sigma}^2$ & 0.000218 & 0.07 & 0.000 & 0.185 \\
$R^2$ & 0.157 &  & 0.153 &  \\
$F$-statistic & 6.767 &  & 6.531 &  \\
\bottomrule
\end{tabular}
\end{table}
\vspace{1em}

\noindent \textbf{Quadratic model}

To capture potential non-linear effects of common players, we extend the model with a quadratic term:
\[
G_i = \beta_0 + \beta_1 C + \beta_2 C^2 + \beta_3 E + \beta_4 V_\mu + \beta_5 \overline{\Sigma}^2 + \varepsilon_i,
\]
which yields the following significant coefficients (Table~\ref{tab:reg_quad5}).

\begin{table}[h!]
\centering
\caption{Quadratic Regression Model for IP One-Third Form Gain}
\label{tab:reg_quad5}
\begin{tabular}{lc}
\toprule
\textbf{Variable} & \textbf{Coefficient (Sig.)} \\
\midrule
(Constant) & -140.401 \quad (0.023) \\
Experts ($E$) & 19.285 \quad (0.000) \\
$V_\mu$ & 0.002 \quad (0.798) \\
$\overline{\Sigma}^2$ & 0.000 \quad (0.144) \\
Common Players ($C$) & 12.375 \quad (0.075) \\
$C^2$ & -0.448 \quad (0.053) \\
\bottomrule
\end{tabular}
\end{table}

The positive linear term and negative quadratic term for $C$ confirm the hypothesized inverted-U relationship:  
moderate overlap among expert selections enhances the potential for skill expression, whereas excessive overlap simplifies decision-making and reduces performance dispersion.

\subsection{Interpretation}

The models collectively indicate:
\begin{enumerate}
    \item \textbf{Expert Density Effect:} Increasing the number of expert teams ($E$) consistently raises the expected gain across all strategies and configurations ($p < 0.001$).
    \item \textbf{Commonality Effect:} The impact of common players ($C$) is non-linear—initially positive, then diminishing, as excessive similarity constrains the range of strategic possibilities.
    \item \textbf{Variability Effects:} Between-team variability ($V_\mu$) aids skill detection by magnifying differences between strategies; within-team variability ($\overline{\Sigma}^2$) contributes less once $V_\mu$ is included.
    \item \textbf{Strategic Classes:} ``Strict on Form'' and ``One-Third'' strategies represent distinct behavioural types rather than scale variations of the same rule.
\end{enumerate}

\subsection{Remarks}

Overall, the model fits are moderate ($R^2 \approx 0.15$--$0.35$), which is reasonable given the inherently stochastic nature of fantasy sports outcomes. This indicates that while meaningful structure exists in the relationships being modeled, a considerable amount of variation is still driven by randomness inherent to real-world sporting events.  

The introduction of the \textbf{Impact Player Substitution} feature amplifies gains across all tested configurations, demonstrating that it is a valid and effective modification aimed at enhancing the expression of skill from the user’s side. This addition allows participants’ decisions to play a more decisive role in determining outcomes, thereby strengthening the skill component within the game’s framework.  

Furthermore, the observed \textbf{commonality effect} assists in designing expert team configurations that enable maximum manifestation of skill within the game, establishing it as a crucial element in overall game design. Another important aspect is the control of \textbf{variability between teams}, which was found to be a significant factor influencing the average gain. Maintaining an appropriate level of variability ensures balanced competition and preserves the sensitivity of outcomes to player skill.

\section{Conclusion}
\label{sec:conclusion}

In this paper, we analysed the limited-pool team selection framework from multiple perspectives to evaluate the presence and visibility of skill in fantasy cricket contests. 
The study combined theoretical simulations, model extensions, and empirical validation using IPL~2024 data to uncover how contest structure and participant choices jointly influence the degree of skill expression. 
Through this multi-stage analysis, we identified key parameters that determine whether outcomes are primarily skill-driven or dominated by chance.

Our simulation analyses provided the first layer of insight into how contest design influences skill visibility. 
In the baseline setup, participants selected from a limited number of expert-designed teams, enabling a controlled examination of outcome variability across different decision strategies. 
Subsequent extensions revealed two distinct pathways for improving skill expression—one from the platform side and another from the participant side. 
From the platform’s perspective, parameters such as the number of available expert teams, the overlap of common players, and the inclusion of an \emph{Impact Player} proved central to shaping the skill dynamics of the contest. 
From the user’s perspective, informed selection among correlated expert teams offered measurable advantages over random choice, confirming that strategic reasoning has a quantifiable impact on performance.

Empirical validation using IPL~2024 data further strengthened these findings. 
The results demonstrated that both the number of experts and the degree of overlap among player selections materially affect the contest’s skill content. 
Specifically, the introduction of the \emph{Impact Player} mechanism increased the differentiation among team outcomes, amplifying the influence of participant decision-making. 
These observations underscore that platform-level design choices can directly affect how much skill the game rewards, thereby offering a pathway to design contests that are demonstrably skill-based.

The regression analyses provided additional quantitative evidence on the structural determinants of skill. 
We observed that the relationship between the number of expert teams and the level of measurable skill is approximately linear—adding more experts consistently increases the potential for skillful decision-making. 
In contrast, the relationship between the number of common players shared across teams and the observed skill follows a non-linear (quadratic) pattern. 
Initially, moderate overlap enhances the visibility of skill by forcing nuanced differentiation among choices, but excessive overlap reduces it by homogenizing outcomes. 
This non-linear pattern captures the trade-off between contest complexity and outcome variance.

Finally, our analyses highlighted that both micro-level variability (differences in individual player performance within a team) and macro-level variability (differences in total team scores across teams) significantly influence skill expression. 

Together, these results reveal that the structure of the contest—particularly the design of expert teams and their statistical relationships—plays a decisive role in determining the extent to which skill governs success. 
This understanding offers practical value to both users, who can refine their strategies, and platform designers, who can optimize contest parameters to promote fairness, engagement, and measurable skill expression.

Overall, this study contributes to the growing literature on quantifying skill in fantasy sports by demonstrating how controlled variations in contest architecture influence the visibility of skill. 
By systematically linking simulation insights with real-world validation, it provides an integrated perspective on how both user strategy and platform design shape the skill–chance balance in fantasy cricket contests.

\begin{appendices}
\section{Concentration Guarantee via Chebyshev's Inequality}
\label{app:dirichlet}
The Dirichlet distribution with parameter vector $\boldsymbol{\alpha} = \alpha \boldsymbol{\pi}$ has the following properties:
\begin{align*}
\mathbb{E}[p_i] &= \pi_i, \\
\mathrm{Var}(p_i) &= \frac{\pi_i(1 - \pi_i)}{\alpha + 1}.
\end{align*}

To ensure that:
\[
\mathbb{P}(|p_i - \pi_i| \geq \beta) \leq \frac{\delta}{n},
\]
for some small $\delta > 0$ and each $i = 1, \dots, n$, we apply \textbf{Chebyshev’s inequality}:
\[
\mathbb{P}(|p_i - \pi_i| \geq \beta) \leq \frac{\mathrm{Var}(p_i)}{\beta^2} = \frac{\pi_i(1 - \pi_i)}{(\alpha + 1)\beta^2} \leq \frac{1}{4(\alpha + 1)\beta^2}.
\]
The upper bound follows from the fact that $\pi_i(1 - \pi_i) \leq \frac{1}{4}$.

To ensure this is at most $\frac{\delta}{n}$ for all $i$, it suffices to have:
\[
\frac{1}{4(\alpha + 1)\beta^2} \leq \frac{\delta}{n} \quad \Rightarrow \quad \alpha \geq \frac{n}{4\delta \beta^2} - 1.
\]

\subsubsection*{Joint Guarantee via Union Bound}

To ensure that \emph{all} components are simultaneously within $\beta$ of their respective true probabilities with probability at least $1 - \delta$, we apply the union bound:
\[
\mathbb{P}\left( \max_{1 \leq i \leq n} |p_i - \pi_i| \geq \beta \right) \leq \sum_{i=1}^n \mathbb{P}(|p_i - \pi_i| \geq \beta) \leq \delta,
\]
provided that each term satisfies:
\[
\mathbb{P}(|p_i - \pi_i| \geq \beta) \leq \frac{\delta}{n}.
\]

Thus, setting
\[
\alpha \geq \frac{n}{4\delta \beta^2} - 1
\]
guarantees that with probability at least $1 - \delta$, the estimated vector $\mathbf{p}$ lies within a $\beta$-ball of $\boldsymbol{\pi}$ in the sup-norm.
\end{appendices}

\end{document}